%% file: manipulation.tex
\documentclass[oneside,letter,12pt,pdflatex]{article}
\usepackage[utf8]{inputenc}
\usepackage[margin=1in]{geometry}
\usepackage{natbib}
\usepackage{booktabs, calc}
\usepackage{amssymb}
\usepackage{amsmath}
\usepackage{bbm}
\usepackage{amsthm} 
\usepackage{threeparttable}
\usepackage{epsfig}
\usepackage{graphics}
\usepackage{lscape}
\usepackage{setspace}
\usepackage{multirow}
\usepackage{tikz}
\usepackage{comment}
\usepackage{subcaption}
\usepackage{algorithm2e}
\usepackage{array}
\usepackage{rotfloat}
\usepackage{units}
\usepackage{textcomp}
\usepackage{mathtools}

\PassOptionsToPackage{normalem}{ulem}
\usepackage{ulem}
\usepackage{microtype}

\newcommand{\bi}{\begin{itemize}}
\newcommand{\ei}{\end{itemize}}
\newcommand{\beq}{\begin{equation}}
\newcommand{\eeq}{\end{equation}}
\newcommand{\be}{\begin{equation}}
\newcommand{\ee}{\end{equation}}

\definecolor{winered}{rgb}{0.5,0,0}

\usepackage[colorlinks=true, linkcolor=winered, citecolor=winered, urlcolor=black, bookmarks=true]{hyperref} 
\usepackage{bookmark}
\begin{document}

\begin{spacing}{1}
\title{Manipulation-Proof Machine Learning\thanks{
We are grateful for helpful conversations with Susan Athey, John Friedman, and Jesse Shapiro. This project would not have been possible without the creative work of Chaning Jang, Simon Muthusi, Nicholas Owsley, and the rest of the team at the Busara Center for Behavioral Economics. We thank numerous audiences for helpful feedback. We are grateful for funding from the Brown University Seed Fund, the Bill and Melinda Gates Foundation, and the Digital Credit Observatory. Bj{\"o}rkegren thanks the W. Glenn Campbell and Rita Ricardo-Campbell National Fellowship at Stanford University, and Microsoft Research for support. This study was pre-registered with the AEA RCT Registry (AEARCTR-0004649), and approved by the IRBs of UC Berkeley, Brown University, and the Kenya Medical Research Institute.}}

\author{
		Daniel Bj{\"o}rkegren\footnote{\small dan@bjorkegren.com} \\ 
			\footnotesize Brown University \ 
		\and
		Joshua E. Blumenstock\footnote{\small jblumenstock@berkeley.edu} \\ 
			\footnotesize U.C. Berkeley \ 
		\and
		Samsun Knight\footnote{\small samsun\_knight@brown.edu}\\ 
	  		\footnotesize Brown University \
	}

\date{\small This version: \today\\ First version: November 30, 2018}
\maketitle

\begin{abstract}
\noindent \small 

An increasing number of decisions are guided by machine learning algorithms.
In many settings, from consumer credit to criminal justice, those
decisions are made by applying an estimator to data on an individual's
observed behavior. But when consequential decisions are encoded in
rules, individuals may strategically alter their behavior to achieve
desired outcomes. This paper develops a new class of estimator that
is stable under manipulation, even when the decision rule is fully
transparent. We explicitly model the costs of manipulating different
behaviors, and identify decision rules that are stable in equilibrium.
Through a large field experiment in Kenya, we show that decision rules
estimated with our strategy-robust method outperform those based on
standard supervised learning approaches.
\end{abstract} 

\vspace{10mm}


\vspace{3mm}
\vspace{2cm}

\end{spacing}

\pagebreak


\section{Introduction}
\label{sec:intro}

An increasing number of important decisions are being made by machine
learning algorithms. Algorithms determine what information we see
online \citep{perlich_machine_2014}; who is hired, fired, and promoted
\citep{brynjolfsson_what_2017}; who gets a loan \citep{bjorkegren_potential_2018},
and whether to give bail and parole \citep{kleinberg_human_2018}.
In the typical machine learning deployment, an individual's observed
behavior is used as input to an estimator that determines future decisions.

These applications of machine intelligence raise two related problems.
First, when algorithms are used to make consequential decisions, they
create incentives for people to reverse engineer or `game.' If agents
understand how their behavior affects decisions, they may alter their
behavior to achieve the outcome they desire.    Second, society
increasingly demands a `right to explanation' about how algorithmic
decisions are made \citep{goodman_european_2016,barocas_fairness_2018}.
For instance, articles 13-15 of the European Union's General Data
Protection Regulation mandate that \textquotedblleft meaningful information
about the logic\textquotedblright{} of automated systems be made available
to data subjects \citep{european_union_eu_2016}. However, such transparency
increases the scope for gaming: the more clearly that agents know
how their behavior affects a decision, the easier it is to manipulate.

These problems result from a simple core. The standard estimators
that are used to construct decision rules assume that the relationship
between the outcome of interest and human behaviors is stable. But
this assumption tends to be violated as soon as a decision rule is
implemented: agents have incentives to change their behavior to achieve
more favored outcomes. When decision rules are gamed, they can produce
decisions that are arbitrarily poor or unsafe. Lenders' portfolios
may be swamped with fraud, social media may be overrun by nefarious
actors, self driving cars can be tricked into crashing \citep{eykholt_robust_2018}.
This problem can undermine the use of machine learning in critical
applications.

There are two common approaches to deal with this problem. The first,
familiar to economists, restricts models to predictors that are presumed
to have a theoretical or structural relationship to the outcome of
interest.\footnote{An extreme version of this restricts to predictors that causally affect
the outcome of interest \citep{kleinberg_how_2018,milli_social_2019}.
This may make manipulation desirable: for example, an exam may induce
students to study and learn general knowledge.} This theory-driven approach amounts to having a dogmatic prior that
the cost of manipulation is either infinite (for included features)
or zero (for excluded features). However, most behaviors are manipulable
at some cost.  The second approach, which we refer to as the `industry
approach', keeps decision rules secret, and periodically updates
the model to account for changes in the relationship between features
and outcomes \citep{bruckner_stackelberg_2011}. However, such `security
through obscurity' exposes current applications to substantial risk
\citeyearpar[NIST][]{national_institute_of_standards_and_technology_guide_2008}.
It also limits the application of machine learning in settings where
secrecy cannot be maintained (e.g., when regulations mandate transparency,
or when consumers learn decision rules directly or through third parties)
or feedback is noisy or delayed (e.g., it may take years for a social
media platform to learn that its content prioritization algorithm
was gamed by foreign actors). There is also no guarantee that the
back and forth between estimation and agents will reach equilibrium.

This paper develops a new approach. We explicitly model the costs
that agents incur to manipulate their behavior, and embed the resulting
game theoretic model within a machine learning estimator. This allows
us to derive estimators that anticipate strategic agents, and which
produce stable decisions even when the decision rule is fully transparent.
We demonstrate, using Monte Carlo simulations, that our `strategy-robust'
estimator performs better than standard models when these costs are
known, even if costs are misspecified. We then test the theory in
a real world environment, through an incentivized field experiment
with 1,557 people in Kenya. We use the experiment to elicit costs
of manipulating behavior, and to show that the strategy-robust approach
leads to more robust machine decisions.

The paper is organized into two main parts. The first part develops
a method to estimate strategy-robust decision rules that are stable
under manipulation. We consider a supervised machine learning framework
for a policymaker making a decision $y_{i}$ for each individual $i$.
Each individual prefers a larger decision $y_{i}$. We observe a \emph{training
}subset of cases that possess both features $\mathbf{x}_{i}$ and
optimal decisions $y_{i}$. The policymaker seeks to estimate a decision
rule $\hat{y}(\mathbf{x}_{i})$ for cases in a \emph{testing} subset
where only features $\mathbf{x}_{i}$ are observed. Standard methods
assume that $\mathbf{x}_{i}$'s are fixed: training and test samples
of $(\mathbf{x}_{i},y_{i})$ are drawn from same distribution. Our
method allows individuals to adjust behavior in response to the incentives
generated by the decision rule: $\mathbf{x}_{i}(\hat{y}(\cdot))$
is a function of the decision rule. As a result, while our training
samples come from an unincentivized distribution $(\mathbf{x}_{i}(0),y_{i})$;
test samples come from $(\mathbf{x}_{i}(\hat{y}(\cdot)),y_{i})$.
We assume individuals pay quadratic costs for manipulating behavior
($\mathbf{x}_{i}$), and that these costs can be parametrized by a
matrix $\mathbf{C}_{i}$. We describe several methods to estimate
this cost matrix, a new object needed to determine how behavior shifts
when incentivized.

To sharpen intuition, we derive results for linear decision rules
of the form $\hat{y}(\mathbf{x})=\boldsymbol{\beta}\mathbf{x}$. 
The resulting estimator takes a simple nonlinear least squares form.
Our method introduces a new notion of fit, which has analogues to
other common linear regression approaches. Ordinary least squares
(OLS) maximizes fit within sample; two stage least squares (2SLS)
sacrifices fit within sample to estimate coefficients that have causal
interpretations; penalized least squares (such as LASSO and ridge)
sacrifice within-sample fit to better generalize to other samples
drawn from the same population. Our method sacrifices fit within sample
to maximize equilibrium fit in the counterfactual where the decision
rule is used to allocate resources, and agents manipulate against
it. Our estimator is an example of a new class of estimator that maximizes
\emph{counterfactual fit}--predictive fit in a counterfactual state
of the world.

We use Monte Carlo simulations to compare this new strategy-robust
approach to common alternatives. OLS can perform extremely poorly
when agents behave strategically. The industry approach, which periodically
retrains the model, can also perform poorly and converge slowly, or
not at all. By contrast, our method adjusts the model to anticipate
manipulation. In simulations where agents respond to the decision
rule, and manipulation costs are known, our approach exceeds the performance
of other estimators. Our approach can exceed the performance of others
even if manipulation costs are misspecified for some cases. Under
certain parameters, the presence of manipulation can \emph{improve}
predictive performance, if it signals unobservables associated with
the outcome of interest \citep[in the spirit of][]{spence_job_1973}.
In these cases, one may wish to use certain features that are manipulable
by the types that you want to screen in, but not by those you want
to screen out. 

In the second part of the paper, we implement and test our method
in the context of a field experiment in Kenya. This experiment allows
us to compare the performance of the strategy-robust estimator to
standard machine learning algorithms in a real-world environment.
Specifically, we built a new smartphone app that passively collects
data on how people use their phones, and disburses monetary rewards
to users based on the data collected. The app is designed to mimic
`digital credit' products that are spreading dramatically through
the developing world \citep{francis_digital_2017}. Digital credit
products similarly collect user data, and convert it into a credit
score using machine learning, based on the insight that historical
patterns of mobile phone use can predict loan repayment \citep{bjorkegren_big_2010,bjorkegren_behavior_2019}.
However, as these systems have scaled, manipulation has become commonplace
as borrowers learn what behaviors will increase their credit limits
\citep{mccaffrey_m-shwari:_2013,bloomberg_phone_2015}.\footnote{A recent survey in Kenya and Tanzania found that one of the top five
reasons people report saving money in digital accounts is to increase
the loan amount qualified for \citep{fsd_kenya_tech-enabled_2018}.} 

This field experiment produces several results. First, consistent
with prior work, we show that a person's mobile phone usage behaviors
($\mathbf{x}_{i}(0)$) can be used to predict characteristics of the
phone user, such as income, intelligence (Raven's matrices), and overall
activity.\footnote{Prior work has used mobile phone data to predict income and wealth
\citep{blumenstock_predicting_2015,blumenstock_estimating_2018},
gender \citep{blumenstock_whos_2010,frias-martinez_gender-centric_2010},
and employment status \citep{sundsoy_estimating_2016}, and loan repayment
\citep{bjorkegren_potential_2018,bjorkegren_behavior_2019}, .} Second, through the use of randomly-assigned experiments, we structurally
estimate $\mathbf{C}_{i}$ in our model, i.e., the relative costs
of manipulating a variety of observed behaviors $\mathbf{x}_{i}$.
Our experiments offer financial incentives to participants for altering
behaviors that are observed through the app, such as increasing the
number of outgoing calls in a given week, or decreasing the number
of incoming text messages. The pattern of costs is intuitive: outgoing
communications are less costly to manipulate than incoming communications;
text messages, which are relatively cheap to send, are more manipulated
than calls, which are relatively expensive. We also find that complex
behaviors (such as the standard deviation of talk time) are less manipulable
than simpler behaviors (such as the average duration of talk time).

The next set of results demonstrate that strategy-robust decision
rules, which account for the costs of manipulation, perform substantially
better than standard machine learning algorithms. We make this comparison
by offering rewards to people who use their phones like a person of
a particular type. For instance, some people receive a message that
says, ``Earn up to 1000 Ksh if the Sensing app guesses that you are
a high income earner, based on how you use your phone,'' while others
receive messages that offer rewards for acting like an ``intelligent''
person, and so forth. Across a variety of such
decision rules, we show that classifications made with the strategy-robust
algorithm are more accurate than classifications from standard algorithms.

Finally, we use our method to estimate the equilibrium cost of algorithmic
transparency, i.e., the cost to the policymaker incurred for disclosing
details of the decision rule. In the experiment, we experimentally
vary the amount of information subjects have about the decision rule
(e.g., the model used to predict the outcome), and show that the relative
performance of the strategy-robust estimator increases with transparency.
While predictive performance decreases by on average 23\% under transparency
for standard machine learning estimators, the strategy-robust estimator
reduces this cost of transparency to approximately 8\%. Overall, this
suggests that the equilibrium cost of moving from a regime where the
decision rules are secret, to one where they are disclosed, to be
less than 8\% in our setting. Our model allows policymakers to bound
this equilibrium cost of transparency even without disclosing decision
rules to the world.

Taken together, the paper develops and tests a new approach to supervised
learning when agents are strategic. This relates to papers from a
variety of sub-literatures have confronted the notion that agents
will act strategically when their actions are used to determine allocations.
Our paper aims to integrate these approaches by applying principles
of mechanism design to the machine learning setting, where data may
have many dimensions and traditional approaches to designing incentive-compatible
allocations are not possible. To our knowledge, this is also the first
paper to estimate and test a strategy-robust machine learning estimator
using data from a field experiment.

\subsection{Connection to Literature}

The dilemma of manipulation is not new. \citet{goodhart_monetary_1975},
in what has since become referred to as `Goodhart's Law', noted that
once a measure becomes a target, it ceases to be a good measure. \citet{lucas_econometric_1976}
also famously observed that historical patterns can warp when economic
policy changes. More broadly, our approach connects with literatures
in both economics and computer science.

Our problem can be viewed as a mechanism design problem. Canonical
signaling models \citep{spence_job_1973} rely on a single crossing
condition to allow full revelation of individual types. In our setting,
like the settings of \citet{frankel_muddled_2019} and \citet{ball_scoring_2019},
there are two forms of heterogeneity: types $\boldsymbol{\theta}_{i}$
and the costs of manipulating behavior $\mathbf{C}_{i}$. \citet{frankel_muddled_2019}
show that unobserved heterogeneity in manipulation costs $\mathbf{C}_{i}$
`muddles' the relationship between behavior $\mathbf{x}_{i}$ and
types $\boldsymbol{\theta}_{i}$, causing the single crossing condition
to fail. That paper shows that muddling reduces the information available
in a market. \citet{ball_scoring_2019} extends that framework to
multiple dimensions of behavior, and in a theoretical model similar
to ours, characterizes and proves the existence of equilibrium. That
paper also considers how the problem is affected by the degree of
commitment available to the policymaker. Relative to this work, our
paper builds a model that can be empirically estimated, which allows
us to probabilistically separate types and costs.\footnote{In a related setting, \citet{hussam_targeting_2017} implement an
incentive compatible mechanism that collects peer reports to estimate
an individual's entrepreneurial ability. That method requires gathering
peer reports from a community during implementation; in contrast,
our approach produces stand in replacements for standard machine learning
models, which can use arbitrary data on behavior. Also related, \citet{holmstrom_moral_1979}
shows that a principal should use any information that has signal
when contracting with an agent. Our method suggests how manipulable
information be downweighted. \citet{eliaz_incentive-compatible_2018}
study a related problem where a ``statistician'' is making decisions
on behalf of an agent, with two-sided incomplete information: the
agent knows his preferred behavior, but the statistician knows the
decision rule. They focus on characterizing incentive-compatible estimators,
and find that commonly-used regularized linear models create incentive
issues.}

Our paper is also related to the problem in public finance of setting
taxes in environments where agents adapt their behaviors. Our method
weights predictors by the inverse of the matrix of the costs of manipulating
them, in a manner similar to \citet{ramsey_contribution_1927}. Relatedly,
\citet{mirrlees_exploration_1971} recommends using proxies when it
is not possible to observe the true income earning ability of potential
beneficiaries. \citet{niehaus_targeting_2013} find that when implementing
agents can be corrupted, considering additional poverty indicators
can worsen the targeting of benefits, by making it more difficult
to verify eligibility. 

Finally, our approach relates to existing strands in the computer
science literature. The theoretical computer science community has
recently considered this problem as one of `strategic classification'
\citep{hardt_strategic_2016,dong_strategic_2018}. This literature
is focused primarily on obtaining computationally efficient learning
algorithms, and how strategic behavior can affect statistical definitions
of fairness \citep{hu_disparate_2019,milli_social_2019}. In computer
security, `adversarial machine learning' considers how strategic adversaries
can systematically undermine supervised learning algorithms, typically
by injecting erroneous data into the model fitting procedure.\footnote{For instance, \citet{bruckner_stackelberg_2011} study adversarial
prediction when the agent acts in response to an observed predictive
model, with an application to spam filtering.  \citet{dong_strategic_2018}
model an iterated industry approach where a policymaker observes how
agents manipulate in response to previous rules, but does not know
their utility functions or costs.} Also related is the concept of `covariate shift', which considers
scenarios where a test distribution differs from the training distribution.
However, it is common to assume that the conditional distribution
$y|\mathbf{x}$ is fixed, and the distribution of $\mathbf{x}$'s
changes exogenously \citep{sayed-mouchaweh_learning_2012}. The manipulation
we consider induces the conditional distribution $y|\mathbf{x}$ to
change endogenously when action is taken based on the estimated relationship.

Thus, papers from a variety of sub-literatures have confronted the
notion that agents will act strategically when their actions are used
to determine allocations. Relative to prior work, our paper makes
two main contributions. First, we develop an equilibrium model of
manipulation that can be estimated using data, which produces a machine
learning estimator that functions well under manipulation even when
the decision rule is fully transparent. And second, to our knowledge
for the first time in any literature, we design and implement a field
experiment that stress-tests such an estimator in a real-world setting
with incentivized agents.

\subsection{Applications and Examples}

Agents game decision rules in a wide variety of empirical settings.
Manipulation has been documented in contexts ranging from New York
high school exit exams \citep{dee_causes_2019} and health provider
report cards \citep{dranove_is_2003}, to pollution monitoring in
China \citep{greenstone_can_2019}, to fish vendors in Chile \citep{gonzalez-lira_slippery_2019}.
In the online advertising industry, firms spend many millions of dollars
each year on search engine optimization, manipulating their websites
in order to receive a higher ranking from search engine algorithms
\citep{borrell_associates_trends_2016}. A quick Google search suggests
over 50 thousand different websites (and 3,000 YouTube videos) contain
the phrase ``hack your credit score.''

We apply our method to an experiment that mimics poverty targeting.
In developing countries, where income is difficult to observe, policymakers
commonly target program eligibility ($y_{i}$) based on easily observable
characteristics or behaviors ($\mathbf{x}_{i}$) \citep{hanna_universal_2018}.
The policymaker may infer a household's type based on the levels of
these variables, or, implicitly, on how they change in response to
incentives.\footnote{Our method thus nests this latter case of self-targeting \citep{nichols_targeting_1982,alatas_self-targeting:_2016},
which identifies beneficiaries based on willingness to engage with
a costly ``ordeal.''} There is evidence that such decision rules induce households to manipulate
their observable features. For instance, \citet{banerjee_lack_2018}
find that adding a question about flat screen TV ownership to a census
caused people to underreport ownership by 16\% on a follow-up survey,
in order to appear less wealthy.\footnote{In other examples from the development literature, \citet{camacho_manipulation_2011}
find that after a program eligibility decision rule was made transparent
to local officials in Colombia, it was manipulated by an amount corresponding
to 7\% of the National Health and Social Security budget. They note,
``there is anecdotal evidence of people moving or hiding their assets,
or of borrowing and lending children.''}

The method we develop is directly relevant to a variety of other settings
where a policymaker derives a decision from a prediction ($y_{i}$)
based on agent behaviors ($\mathbf{x}_{i}$). These include other
supervised settings where it is possible to obtain a ground truth
value of $y_{i}$ for a training sample of individuals. For instance, in credit scoring
applications, a decision about whether it is prudent to provide a
loan ($y_{i}$) is made based on characteristics on the potential
borrower (traditional credit scores are based on the borrower's formal
credit history, but increasingly the characteristics $\mathbf{x}_{i}$
include private data like mobile phone usage \citep{bjorkegren_behavior_2019}
and social network structure \citep{wei_credit_2015}). It also includes
settings where no definite ground truth of $y_{i}$ exists. Search
engines, social media, and spam filters attempt to determine the quality
of a piece of content ($y_{i}$) based on features that can be observed
($\mathbf{x}_{i}$: keywords, reputation of the sender, inbound links).
Manipulating these features may be costly directly, or may undermine
the author's intent in distributing the content. Similarly, `report
cards' for universities, hospitals, and doctors attempt to determine
quality ($y_{i}$) based on indicators ($\mathbf{x}_{i}$: alumni
giving rates, endowment size, acceptance rates, graduation rates).\footnote{Our model does not consider behaviors $\mathbf{x}_{i}$ that have
a causal relationship to $y_{i}$, where manipulation can be productive
\citep{kleinberg_how_2019}. It thus would not cover report card variables
that directly influence quality, nor the case of a student who `games'
a test by studying ($\mathbf{x}_{i}\uparrow$), and as a result improves
their knowledge ($y_{i}\uparrow$). The approach could be extended
to cover such cases.}

The remainder of the paper is organized as follows. The next section
introduces our theory. Section \ref{sec:estimation} describes estimation.
Section \ref{sec:experiment} describes the results of our field experiment.
Section \ref{sec:extensions} discusses extensions. Section \ref{sec:conclusion}
concludes.

\section{Theory}
\label{sec:model}

This section introduces the model underlying our estimator, and demonstrates
the intuition with simulations.

\subsection{Model}

A policymaker observes a \emph{training }subset of cases that possess
both features $\mathbf{x}_{i}$ and optimal decisions $y_{i}$.
The policymaker also obtains information on the costs of manipulating
features, which will be detailed later. The policymaker would like
to estimate the parameters of a decision rule $\hat{y}(\mathbf{x}_{i})$
for cases in a \emph{testing} subset where only features $\mathbf{x}_{i}$
are observed, and may be manipulated.

A policymaker has a preferred action $y_{i}$ for each \textbf{\emph{individual}}
$i$, denominated in units of individuals' utility. The action $y_{i}$
can be projected onto $i's$ bliss behavior $\boldsymbol{\text{\ensuremath{\underbar{x}}}}_{i}$
by the equation $y_{i}=b_{0}+\boldsymbol{b}'\boldsymbol{\text{\ensuremath{\underbar{x}}}}_{i}+e_{i}$,
with $e_{i}\perp\boldsymbol{\text{\ensuremath{\underbar{x}}}}_{i}$
representing idiosyncratic preference.

However, the policymaker observes an individual's actual behavior
$\mathbf{x}_{i}$, which may differ from their bliss level $\boldsymbol{\text{\ensuremath{\underbar{x}}}}_{i}$.
It selects a deterministic decision rule of the form\footnote{Although randomizing a decision rule may make it harder to manipulate,
it undermines a major goal of transparency: that people know how they
are evaluated.}: 
\begin{equation*}
\hat{y}(\mathbf{x}_{i})=\beta_{0}+\boldsymbol{\beta}'\mathbf{x}_{i}
\end{equation*}

Individuals can manipulate their behavior $\mathbf{x}_{i}$ away from
their bliss level $\boldsymbol{\text{\ensuremath{\underbar{x}}}}_{i}$
at some cost. $i$ earns utility from the decision minus these costs:

\begin{equation*}
u_{i}=\hat{y}(\mathbf{x}_{i})-c(\mathbf{x}_{i},\boldsymbol{\text{\ensuremath{\underbar{x}}}}_{i})
\end{equation*}

For simplicity, we consider the case where the utility from the decision
exactly coincides with the policymaker's prediction.\footnote{That is, we consider the case where the utility of the decision $u(\hat{y})=\hat{y}$,
which holds in our experiment. Under more general functions $u(\cdot)$,
our model would represent a linear approximation. One could easily
generalize our framework to allow for more general functional forms.}

Individuals $i$ are heterogeneous in two respects, bliss behaviors
$\boldsymbol{\text{\ensuremath{\underbar{x}}}}_{i}$ and gaming ability
$\gamma_{i}$ (as in \citet{frankel_muddled_2019}).

Manipulation costs are quadratic:

\begin{equation*}
c(\mathbf{x}_{i},\boldsymbol{\text{\ensuremath{\underbar{x}}}}_{i})=\frac{1}{2}(\mathbf{x}_{i}-\boldsymbol{\text{\ensuremath{\underbar{x}}}}_{i})'C_{i}(\mathbf{x}_{i}-\boldsymbol{\text{\ensuremath{\underbar{x}}}}_{i})
\end{equation*}

for matrix $C_{i}$:

\begin{equation*}
C_{i}=\frac{1}{\gamma_{i}}\begin{bmatrix}\alpha_{11} & \cdots & \alpha_{K1}\\
\vdots & \ddots & \vdots\\
\alpha_{1K} & \cdots & \alpha_{KK}
\end{bmatrix}
\end{equation*}

Different behaviors may be differentially hard to manipulate, by themselves
(the diagonal $\alpha_{kk}$) or in conjunction with other behaviors
(the off diagonals $\alpha_{kj}$). And different people may find
it easier or harder to manipulate ($\gamma_{i}$): for example, people
with more technical savvy or lower opportunity cost of time may find
it easier to game decision rules.

When $i$ knows the decision rule $\hat{y}(\mathbf{x}_{i})$ and receives
benefits according to it, he will optimally manipulate behavior to
level:

\begin{equation*}
\mathbf{x}_{i}^{*}(\boldsymbol{\beta})=\boldsymbol{\text{\ensuremath{\underbar{x}}}}_{i}{\color{blue}+C_{i}^{-1}\boldsymbol{\beta}}
\end{equation*}

When behavior is not incentivized ($\boldsymbol{\beta}=\boldsymbol{0}$),
optimal behavior equals the bliss level ($\mathbf{x}_{i}^{*}(\boldsymbol{0})=\boldsymbol{\text{\ensuremath{\underbar{x}}}}_{i}$).
However, as $\boldsymbol{\beta}$ moves away from zero, behavior moves
in the same direction, downweighted by the cost of manipulation (as
highlighted in blue). 

\emph{Decision rules.} The policymaker faces expected squared loss:

\begin{equation*}
L\left(\hat{y}(\cdot)\right)=E_{i}\left[\left[y_{i}-\hat{y}\left(\mathbf{x}_{i}\left(\hat{y}(\cdot)\right)\right)\right]^{2}+M(\cdot)\right]
\end{equation*}

The first term represents fit of the model \emph{in the counterfactual}
where the model is implemented and agents manipulate behavior. If
the policymaker additionally cares about the costs that individuals
incur manipulating, this manipulation cost results in additional term
$M(\cdot)$.

Our \textbf{strategy-robust decision rule} is given by:

\begin{equation}
\boldsymbol{\beta}^{stable}=\arg\min_{\boldsymbol{\beta}}\left(\frac{1}{N}\sum_{i}\left[y_{i}-\beta_{0}-\boldsymbol{\beta}'(\boldsymbol{\text{\ensuremath{\underbar{x}}}}_{i}{\color{blue}+C_{i}^{-1}\boldsymbol{\beta}})\right]^{2}+\ldots\right)\label{eq:estimator}
\end{equation}

which deviates from ordinary least squares due to the term ${\color{blue}C_{i}^{-1}\boldsymbol{\beta}}$
which captures manipulation in response to $\boldsymbol{\beta}$.
Additional terms `$\ldots$' can include any weight $M(\cdot)$ the
policymaker places on manipulation costs incurred by agents, and any
regularization terms $R_{\lambda^{decision}}(\cdot)$.

\subsubsection*{Discussion}

If the policymaker only cares about targeting performance ($M(\cdot)\equiv0$)
and there are no additional regularization terms ($R(\cdot)\equiv0$),
then ours is a nonlinear least squares estimator. Moment conditions
are given by:%
\begin{equation*}
E\left[\boldsymbol{\text{\ensuremath{\underbar{x}}}}_{i}\cdot\left(y_{i}-\beta_{0}-\boldsymbol{\beta}'(\boldsymbol{\text{\ensuremath{\underbar{x}}}}_{i}+C_{i}^{-1}\boldsymbol{\beta})\right)\right]=-2E\left[C_{i}^{-1}\boldsymbol{\beta}\cdot\left(y_{i}-\beta_{0}-\boldsymbol{\beta}'(\boldsymbol{\text{\ensuremath{\underbar{x}}}}_{i}+C_{i}^{-1}\boldsymbol{\beta})\right)\right]
\end{equation*}

This suggests that the estimator imposes that equilibrium errors in
the counterfactual are \emph{less }than orthogonal to individual types
$\boldsymbol{\text{\ensuremath{\underbar{x}}}}_{i}$: they equal the
negative of an adjustment factor $2C_{i}^{-1}\boldsymbol{\beta}$
that accounts for the fact that $\boldsymbol{\beta}$ induces a marginal
incentive to respond. When $C_{i}\equiv\boldsymbol{\infty}$, the
resulting estimator corresponds to OLS.

When the policymaker cares about not only the resulting allocation,
but also the manipulation costs that individuals incur, this is accompanied
by the term $M(\cdot)$, which can take a different form depending
on policymaker preferences. An entity that is narrowly concerned with
its own objective (e.g., profits in the case of a firm) may thus select
different decision rules from those that maximize social welfare (for
example, a firm may be satisfied with an equilibrium where all individuals
expend welfare gaming a test, where a social planner may not).\footnote{For example, the policymaker may place weight $w$ on the sum of manipulation
costs: $M(\cdot)=w\sum_{i}c(\boldsymbol{\text{\ensuremath{\underbar{x}}}}_{i}+C_{i}^{-1}\boldsymbol{\beta},\boldsymbol{\text{\ensuremath{\underbar{x}}}}_{i})$.
The Supplemental Appendix derives a microfounded term for the case
of proxy means testing.}

To reduce overfitting in small samples, one may also include common
forms of regularization; for example,  $R_{\lambda^{decision}}^{LASSO}(\boldsymbol{\boldsymbol{\beta}})=\lambda^{decision}\sum_{k>0}\left|\beta_{k}\right|$
or $R_{\lambda^{decision}}^{ridge}(\boldsymbol{\boldsymbol{\beta}})=\lambda^{decision}\sum_{k>0}\beta_{k}^{2}$.
Hyperparameter $\lambda^{decision}$ can be set with cross validation
in the baseline sample. Under these regularization terms, when $M(\cdot)\equiv0$
and $C_{i}\equiv\boldsymbol{\infty}$ the resulting estimator corresponds
to LASSO, or ridge, respectively.

\subsection{Intuition\label{sec:Intuition}}

We demonstrate the method with Monte Carlo simulations.

We derive desired payments $\mathbf{y}$, from individual types $\boldsymbol{\text{\ensuremath{\underbar{x}}}}$
and payment rule $\mathbf{b}$, with deviations $\boldsymbol{e}$.
We then assess decision rules $\hat{y}(\mathbf{x})$ based on observed
behaviors $\mathbf{x}$ generated with different estimators. Our strategy-robust
estimators anticipate that behaviors $\mathbf{x}$ may change when
they are used in a decision rule, factoring in manipulation costs
$\mathbf{C}$. This section assumes that manipulation costs are known.

\subsubsection*{Comparative Statics}

We consider a case where $x_{1}$ is more predictive than $x_{2}$
in baseline behavior, but would be easily manipulated if used in a
decision rule ($b_{1}>b_{2}$ but $\alpha_{11}\ll\alpha_{22}$).

\begin{figure}
\caption{Common vs. Strategy Robust Estimators \label{fig:Comparative-Statics-Penalized}}
\begin{centering}
\begin{tabular}{>{\raggedright}p{5.5cm}>{\raggedright}p{5cm}>{\raggedright}p{5.5cm}}
\multicolumn{1}{c}{(a) $\boldsymbol{\beta}^{OLS}(\nicefrac{1}{\gamma})$} & \multicolumn{1}{c}{(b) $\boldsymbol{\beta}^{LASSO}(\lambda;\gamma=1)$} & \multicolumn{1}{c}{(c) $\boldsymbol{\beta}^{stable}(\nicefrac{1}{\gamma})$}\tabularnewline
\includegraphics[height=4cm]{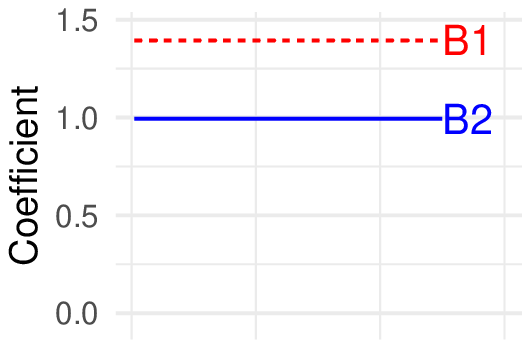} &
 \includegraphics[height=4cm]{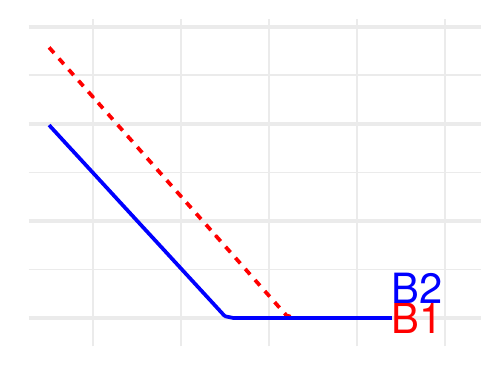} &
 \includegraphics[height=4cm]{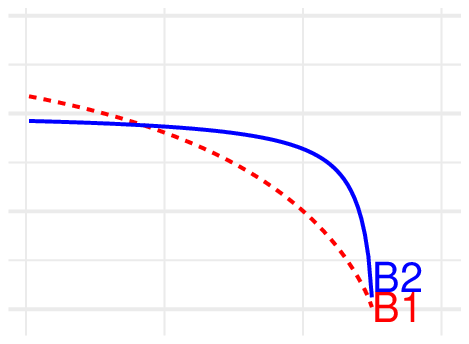}\tabularnewline
\includegraphics[height=2.66cm]{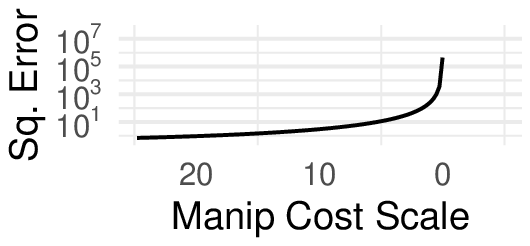} & 
\includegraphics[height=2.66cm]{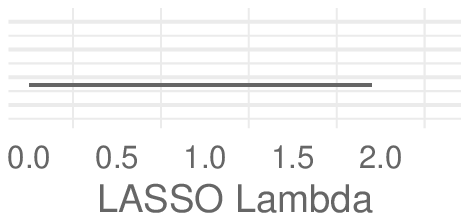} & 
\includegraphics[height=2.66cm]{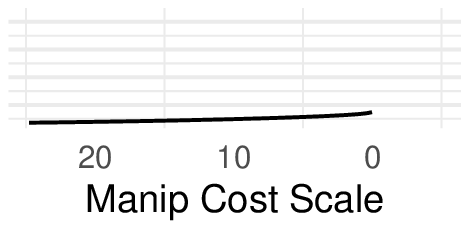}\tabularnewline
 & \centering{}{\scriptsize{}Manipulation Cost = 1} & \tabularnewline
 && \\
\multicolumn{3}{>{\raggedright}p{15cm}}{\footnotesize{\emph{Note: } The first behavior is more predictive ($b_{1}>b_{2}$), but is easily manipulable ($\alpha_{11}\ll\alpha_{22}$). \textbf{(a)} OLS performance deteriorates substantially when behavior can be manipulated. (\textbf{b}) LASSO penalization favors $x_{1}$, which will be manipulated as soon as the decision rule is implemented. \textbf{(c)} Our method anticipates that $x_{1}$ will be manipulated if it is incentivized. It shifts weight to $x_{2}$ as behavior becomes manipulable.}} \tabularnewline 
\\ [-1.5ex] \hline \\ [-1ex]
\multicolumn{3}{>{\raggedright}p{15cm}}{\footnotesize{$\boldsymbol{\text{\ensuremath{\underbar{x}}}}_{i}\overset{iid}{\sim}N\left(0,1\right)$,
$\mathbf{b}=[1.4,1]$, $\mathbf{C}^{het}=\frac{1}{\gamma\gamma_{i}}\left[\begin{array}{cc}
4 & 0\\
0 & 32
\end{array}\right]$, $\frac{1}{\gamma_{i}^{het}}\overset{iid}{\sim}Uniform\left[0,10\right]$,
$e_{i}\overset{iid}{\sim}N\left(0,0.25\right)$. Squared error measured
on an out of sample draw from the same population, incentivized to
that decision rule.}

}\tabularnewline
\end{tabular}
\par\end{centering}
\end{figure}

Figure \ref{fig:Comparative-Statics-Penalized} compares our method
to OLS and LASSO, which mistakenly place most weight on $x_{1}$.
OLS maximizes predicted performance within the unincentivized sample
$(\mathbf{x}_{i}(\boldsymbol{0}),y_{i})$; as shown in Figure \ref{fig:Comparative-Statics-Penalized}a,
it performs poorly as manipulation becomes easier. Figure \ref{fig:Comparative-Statics-Penalized}b
shows that for a given cost of manipulation, LASSO shrinks these coefficients.
However, when LASSO selects variables, it does exactly the wrong thing:
it kicks $x_{2}$ out of the regression first. In contrast, our method
considers how predictive features will be in equilibrium when the
decision rule is implemented: $(\mathbf{x}_{i}(\boldsymbol{\beta}),y_{i})$.
As shown in Figure \ref{fig:Comparative-Statics-Penalized}c, when
manipulation costs are high, our method approaches OLS, but as manipulation
becomes easier, our method substantially penalizes $x_{1}$. Our method
can also be combined with LASSO or ridge penalization to fine tune
out of sample fit.\footnote{See Appendix Figure \ref{fig:Comparative-Statics-Additional} for
a comparison to ridge regression, as well as a demonstration of combining
our method with ridge penalization.}

If each feature is equally costly to manipulate, our method shrinks
them together, similar to ridge regression, as shown in Figure \ref{fig:Comparative-Statics-Additional}.
If all individuals have the same gaming ability ($\gamma_{i}\equiv\gamma$),
then manipulation shifts behavior uniformly and does not affect predictive
performance. However, even though predictive performance is high,
individuals' can spend substantial utility on manipulation. Figure \ref{fig:Comparative-Statics-Simple} develops this intuition further, 
by showing how the strategy-robust method penalizes indicators that are easy to shift:
Figure \ref{fig:Comparative-Statics-Simple}a shows the effect of
scaling the cost of one behavior ($x_{2}$). As the cost of manipulating
that particular behavior ($\alpha_{22}$) decreases, it is penalized,
and weight is shifted to other behaviors. The method also penalizes
indicators that make it easier to shift other predictive indicators
(in a manner similar to \citet{ramsey_contribution_1927} taxation).
Figure \ref{fig:Comparative-Statics-Simple}b shows that the effect
of cost interactions: when manipulating $x_{1}$ makes it easier to
manipulate $x_{2}$ ($\alpha_{12}$ sufficiently negative), our method
further reduces weight on $x_{1}$ .


\subsubsection*{Performance}

Table \ref{tab:Monte-Carlo-Simulation-NonEquilibrating} shows the
results of an example Monte Carlo simulation, chosen to demonstrate
how standard approaches can fail. In this simulation, type $\text{\ensuremath{\underbar{x}}}_{1}$
has a large weight in the desired payment ($b_{1}=3$) relative to
the other two dimensions ($b_{2}=b_{3}=0.1$); however, the resulting
behavior $x_{1}$ is much easier to manipulate ($\alpha_{11}=1$ vs.
$\alpha_{22}=2$ and $\alpha_{33}=4$).

In this environment, OLS considers the static relationship in the
unmanipulated data. This rule would perform well if behavior were
held fixed (no manipulation column); however, once consumers adjust
to the rule, it makes terrible decisions (manipulation column).

The industry approach would retrain (refresh) this model after this
manipulation. If we observe how consumers adjust their behavior and
reestimate OLS, we obtain $\boldsymbol{\beta}^{OLS(1)}$, which places
negative weight on the manipulated $x_{1}$. However, its also makes
terrible decisions when consumers respond to it. We can try to do
better by repeatedly allowing individuals to best respond, and then
reestimating the decision rule. But even with perfect information
and no changes in the environment, this process can make poor decisions
en route to convergence, or may not converge at all. If we estimate
$\boldsymbol{\beta}^{OLS(r)}$ using data from all prior periods ($1,\ldots,r-1$),
it continues to make terrible decisions over several iterations of
the algorithm designer announcing decision rules to consumers, and
learning from how they respond. While the performance of these decisions
then begins to improve, it would require sequentially announcing over
a thousand different rules to consumers, and learning from how they
responded to each one, to approach equilibrium. (See the second set
of estimates in Table \ref{tab:Monte-Carlo-Simulation-NonEquilibrating}.)
If we instead rely on only recent data, estimating $\boldsymbol{\beta}^{OLS(r)}$
using only data from the prior period ($r-1$), this approach does
not reach equilibrium: it alternates between decision rules that place
high and low weight on $x_{1}$ (see Table \ref{tab:Monte-Carlo-Simulation-NonEquilibratingNonCum}).
Thus standard approaches can perform poorly even in ideal cases.
If there were noise or frictions in learning, the risks of this approach
are greater: the rule may appear to be performing well, and suddenly
be devastatingly undermined (for example, \citet{gonzalez-lira_slippery_2019}
find that increased enforcement of a ban on selling an endangered
fish can lead vendors to learn about the decision rule, and more effectively
undermine it).

In contrast, our strategy-robust estimator ($\boldsymbol{\beta}^{stable}$)
anticipates that including a behavior in the decision rule will shift
that behavior. It penalizes the easily manipulable behavior $x_{1}$,
and shifts weight to behaviors that are harder to manipulate ($x_{2}$
and $x_{3}$). It sacrifices performance in the environment in which
it is trained (in sample, no manipulation) for performance in the
counterfactual where there is manipulation. When individuals manipulate
as described in the model, our estimator exceeds the performance of
other estimators.

Our method can reduce risk even if manipulation costs are misestimated.
We consider a case with two measurement mistakes: (a) all off diagonal
elements are set to zero, and (b) the estimated costs of manipulation
are two times too large. Performance deteriorates relative to the
case where we know the true cost matrix, but our method still outperforms
OLS in the presence of manipulation. One can use our method as a first
step towards equilibrium, and then follow it with the industry approach;
as shown in the bottom rows, doing so skips the terrible decisions
made in the first two iterations of the industry approach.

\input{tables/mc-basic}

\subsubsection*{Manipulation can improve performance}

Manipulation can \emph{improve} performance, if ease of manipulation
($\gamma_{i}$) is correlated with the outcome ($y_{i}$). In that
case, manipulation itself represents a signal of the underlying type,
as in \citet{spence_job_1973}, and applications of self-targeting
\citep{nichols_targeting_1982,alatas_self-targeting:_2016}. An example
is shown in Table \ref{tab:Monte-Carlo-Simulation-ManipulationSignal}:
manipulation improves the performance even of na{\"i}ve estimators, as
shown in the first two rows. Our method can additionally exploit cost
heterogeneity, and thus further improves performance as shown in the
third row.

\section{Estimation}
\label{sec:estimation}

Our model can be fully estimated with experimental data. To estimate
manipulation costs, we hire study participants to undermine component
parts of the model, and gauge how sensitive these manipulations are
to incentives.

We observe multiple time periods. Each period, an individual may desire
to deviate from bliss behavior due to manipulation, or shocks that
are common ($\boldsymbol{\mu}_{t}$) or individual specific ($\boldsymbol{\epsilon}_{it}$):
\[
u_{it}=\hat{y}(\mathbf{x}_{i})-c(\mathbf{x}_{i},\boldsymbol{\text{\ensuremath{\underbar{x}}}}_{i})+(\boldsymbol{\mu}_{t}+\boldsymbol{\epsilon}_{it})\cdot(\mathbf{x}_{i}-\text{\ensuremath{\underbar{\ensuremath{\mathbf{x}}}}}_{i})
\]

where both components are mean zero: $E\boldsymbol{\mu}_{t}=\boldsymbol{0}$
and $E\boldsymbol{\epsilon}_{it}=\boldsymbol{0}$. Then, in week $t$
we will observe behavior:

\begin{equation}
\mathbf{x}_{it}^{*}(\boldsymbol{\beta})=\boldsymbol{\text{\ensuremath{\underbar{x}}}}_{i}+\boldsymbol{\mu}_{t}+\boldsymbol{\epsilon}_{it}+C_{i}^{-1}\boldsymbol{\beta}\label{eq:x(beta)}
\end{equation}

We parameterize the inverse of the cost matrix as follows:

\[
C_{i}^{-1}=\gamma_{i}\cdot C^{-1}
\]

with elements of inverse costs defined for convenience as:

\[
C^{-1}\eqqcolon\begin{bmatrix}c_{11} & \cdots & c_{K1}\\
\vdots & \ddots & \vdots\\
c_{1K} & \cdots & c_{KK}
\end{bmatrix}
\]

Gaming ability includes two types of heterogeneity:

\[
\gamma_{i}=e^{-\boldsymbol{\omega}\mathbf{z}_{i}}+v_{i}
\]

It is allowed to vary with characteristics $\mathbf{z}_{i}$ that
are observable in the training sample (but need not be observed in
an implementation sample; for example, we survey participants on tech
savviness). It also includes unobserved heterogeneity $v_{i}\sim V$
with $Ev_{i}=0$, which will enter the model as random effects.

We estimate strategy-robust decision rules in two steps.

\subsection{Primitives\label{subsec:Manipulation-Costs}}

We first estimate primitives: types $\boldsymbol{\text{\ensuremath{\underbar{x}}}}$,
cost parameters $\boldsymbol{\omega}$ and $C^{-1}$, and the distribution
of unobserved gaming ability $V$.

\subsubsection*{Types}

We infer types $\boldsymbol{\text{\ensuremath{\underbar{x}}}}$ by
observing baseline behavior prior to the implementation of a decision
rule. When $\boldsymbol{\beta}=\boldsymbol{0}$, behavior will not
be manipulated. We can estimate types and time period fixed effects
with moment conditions derived from the equation:

\begin{equation}
\mathbf{x}_{it}^{*}(\boldsymbol{0})=\boldsymbol{\text{\ensuremath{\underbar{x}}}}_{i}+\boldsymbol{\mu}_{t}+\boldsymbol{\epsilon}_{it}\label{eq:TypeEst}
\end{equation}

including only time periods where $\boldsymbol{\beta}=\boldsymbol{0}$.

\subsubsection*{Costs}

Our main specification recovers manipulation costs experimentally.
Each week we randomly assign individuals to a decision rule $\boldsymbol{\beta}_{it}$.
The decision rule may be a control, in which case $\boldsymbol{\beta}_{it}\equiv0$.
Or, it may be a treatment group that incentivizes one behavior $k\in1...K$,
by disclosing a rule that pays incentives for $k$: $\beta_{itk}>0$
but not for other behaviors: $\beta_{itj}=0$ for $j\ne k$. These
treatments make it possible to recover the inverse cost matrix (diagonal
and off-diagonal elements), as well as heterogeneous gaming ability
(observed $\boldsymbol{\omega}$ and unobserved $V$).

\subsubsection*{Moment Conditions}

We recover all parameters jointly with the following moment conditions.

Incentives are orthogonal to idiosyncratic behavior shocks ($E[\boldsymbol{\beta}_{itk}\epsilon_{itj}]=0$).
For each pair of behaviors $jk$ (including $j=k$) this yields sample
moment condition:

\[
0=\frac{1}{N}\sum_{i=1}^{N}\beta_{itk}\left[x_{ijt}-\text{\ensuremath{\underbar{x}}}_{ij}-\mu_{jt}-\beta_{itj}\left(e^{-\boldsymbol{\omega}\mathbf{z}_{i}}\cdot c_{kj}\right)\right]
\]

We also have $E[\epsilon_{itj}]=0$: for each time period $t$ and
behavior $k$, we obtain:

\[
\mu_{kt}=\frac{1}{N}\sum_{i=1}^{N}\left[x_{ikt}-\text{\ensuremath{\underbar{x}}}_{ik}-\boldsymbol{\beta}_{it}\left(e^{-\boldsymbol{\omega}\mathbf{z}}\cdot C^{-1}\right)\right]
\]

For each individual $i$ and behavior $k$, we obtain:

\[
\text{\ensuremath{\underbar{x}}}_{ik}=\frac{1}{T}\sum_{t=1}^{T}\left[x_{ikt}-\mu_{kt}-\boldsymbol{\beta}_{it}\left(e^{-\boldsymbol{\omega}\mathbf{z}}\cdot C^{-1}\right)\right]
\]

given $T$ observations.

Unobserved heterogeneity is mean zero ($E[v_{i}]=0$), yielding:

\[
0=\frac{1}{T}\sum_{i,k,t\textnormal{ where \ensuremath{k} incentivized}}\left[\frac{x_{ikt}-\text{\ensuremath{\underbar{x}}}_{ik}-\mu_{kt}}{C^{-1}\boldsymbol{\beta}_{it}}-e^{-\boldsymbol{\omega}\mathbf{z}_{i}}\right]
\]

Each heterogeneity characteristic $z\in\mathbf{z}$ is orthogonal
to unobserved heterogeneity ($E[z_{i}v_{i}]=0$), yielding:

\[
0=\frac{1}{T_{z}}\sum_{i}\text{\ensuremath{z}}_{i}\sum_{k,t\textnormal{ where \ensuremath{k} incentivized}}\left[\frac{x_{ikt}-\text{\ensuremath{\underbar{x}}}_{ik}-\mu_{kt}}{C^{-1}\boldsymbol{\beta}_{it}}-e^{-\boldsymbol{\omega}\mathbf{z}_{i}}\right]
\]

These moment conditions jointly identify $\boldsymbol{\text{\ensuremath{\underbar{x}}}}$,
$C^{-1}$, and $\boldsymbol{\omega}$.

\subsubsection*{Joint Estimation}

We jointly solve for the parameters to minimize the squared distance
from zero:

\[
L(\boldsymbol{\text{\ensuremath{\underbar{x}}}},C^{-1},\boldsymbol{\boldsymbol{\omega}})+R_{costs}^{\boldsymbol{\lambda}^{costs}}(C^{-1},\boldsymbol{\boldsymbol{\omega}})
\]

where $L(\cdot)$ represents the associated general method of moments
(GMM) loss function.

\subparagraph*{\emph{Penalization and Cross Validation}}

We make include two adjustments to reduce overfitting of the cost
matrix to our limited dataset. First, we impose the constraint that
incentivizing a behavior increases it: $c_{jj}>0$. Second, we regularize
the cost estimates:

\[
R_{costs}^{\boldsymbol{\lambda}^{costs}}(\cdot)=\left[\lambda_{diagonal}^{costs}\sum_{k}c_{kk}^{2}+\lambda_{offdiagonal}^{costs}\sum_{j\ne k}c_{jk}^{2}\right]\left[\sum_{i}e^{-2\boldsymbol{\omega}\mathbf{z}}\right]
\]

where we allow the possibility of using separate hyperparameters $\boldsymbol{\lambda}^{costs}=\{\lambda_{diagonal}^{costs},\lambda_{offdiagonal}^{costs}\}$
for diagonal and off diagonal costs. These penalize the cost of manipulation
towards infinity (ease of manipulation towards zero), which will tend
to penalize our method's estimates towards standard methods (OLS/LASSO/etc).

We jointly solve for parameters $\boldsymbol{\text{\ensuremath{\underbar{x}}}}$,
$C^{-1}$, and $\boldsymbol{\omega}$, and hyperparameters $\boldsymbol{\lambda}^{costs}$
to minimize out of sample prediction error, using cross validation.
Then, we impose the optimal $\boldsymbol{\lambda}^{costs}$ and jointly
estimate $\boldsymbol{\text{\ensuremath{\underbar{x}}}}$, $C^{-1}$,
and $\boldsymbol{\omega}$ on the full sample.

\subsubsection*{Unobserved Gaming Ability}

After estimating these parameters, we back out the distribution of
unobserved gaming ability $V$ in two steps. First we compute whether
each individual manipulates more or less than predicted during incentivized
weeks:

\emph{
\[
\tilde{v}_{i}=\frac{1}{K_{i}}\sum_{k}\frac{1}{T_{i}}\sum_{t\textnormal{ where \ensuremath{k} incentivized}}\left[\frac{x_{ikt}-\text{\ensuremath{\underbar{x}}}_{ik}-\mu_{kt}}{C^{-1}\boldsymbol{\beta}_{it}}-e^{-\boldsymbol{\omega}\mathbf{z}_{i}}\right]
\]
}

Second, to reduce the impact of noise and outliers, we shrink and
winsorize these backed out shocks. We form the empirical distribution
$V=\{\max(\phi\cdot\tilde{v}_{i},\text{\ensuremath{\underbar{v}}})\}_{i}$,
where $\text{\ensuremath{\underbar{v}}}$ is the lowest value of $\tilde{v}$
that leads to a nonnegative implied gaming ability.\footnote{That is, $\text{\ensuremath{\underbar{v}}}=\min_{i}(\tilde{v_{i}}|\tilde{v}_{i}\ge\min_{j}(e^{-\boldsymbol{\omega}\mathbf{z}_{j}}))$.}
We set the shrinkage factor $\phi$ to 0.005 so that less than 5\%
of distribution is winsorized.\footnote{After shrinkage, 4.1\% of observations are winsorized.}
This yields a distribution of costs $C_{i}$.

\subsection{Decision Rules}

Given these primitives, a strategy robust decision rule is given by:

\[
\boldsymbol{\beta}^{stable}=\arg\min_{\boldsymbol{\beta}}E\left[\frac{1}{N}\sum_{i}\left[y_{i}-\beta_{0}-\boldsymbol{\beta}'(\boldsymbol{\text{\ensuremath{\underbar{x}}}}_{i}+C_{i}^{-1}\boldsymbol{\beta})\right]^{2}+R_{decision}^{\lambda^{decision}}(\boldsymbol{\beta},\mathbf{y},\mathbf{C})\right]
\]

taken over expectation over $C_{i}$, and given decision rule regularization
term $R^{\lambda^{decision}}(\cdot)$. Hyperparameter $\lambda^{decision}$
is set through cross validation in the unmanipulated sample (where
we can observe ground truth):

\[
\lambda^{decision}=c.v.\arg\min_{\lambda^{cv}}\left[\min_{\boldsymbol{\beta}^{naive}}\left[\frac{1}{N}\sum_{i}\left[y_{i}-\beta_{0}^{naive}-\boldsymbol{\beta}^{naive}\boldsymbol{\text{\ensuremath{\underbar{x}}}}_{i}\right]^{2}+R_{decision}^{\lambda^{cv}}(\boldsymbol{\beta}^{naive},\mathbf{y},\mathbf{C})\right]\right]
\]

\section{Experiment}
\label{sec:experiment}

We designed a field experiment to test the performance of our strategy-robust
estimator in a real-world setting. Design started in 2017. Working
with the Busara Center for Behavioral Economics in Nairobi, we developed
and deployed a new smartphone-based application (`app') to 1,557 research
subjects. The app was designed to mimic the key features of the `digital
credit' apps that are quickly transforming consumer credit in developing
countries \citep{francis_digital_2017}. In Kenya, at the time of
our study, \citet{cgap_kenyas_2018} estimates that 27\% of all adults
had an outstanding `digital credit' loan. These phone-based apps construct
an alternative credit score ($\hat{y_{i}}$) based on how each applicant
uses their phone ($\boldsymbol{x}_{i}$; \citet{bjorkegren_big_2010,bjorkegren_behavior_2019}).
The app we built similarly collects data on how each subject uses
their phone, and uses that data to make cash transfer decisions. This
section describes the app and experimental design (Section \ref{sec:AppDesign});
estimates costs of manipulation and derives strategy-robust decision
rules using our method; and compares the performance of these new
estimators to traditional learning algorithms (Section \ref{sec:results}).
Our design was pre-specified in a pre-analysis plan registered in
the AEA RCT registry under AEARCTR-0004649.

\subsection{Experimental design and smartphone app \label{sec:AppDesign}}

Our experiment is intended to create an environment with incentives
similar to those of a `digital credit' lending app. These apps run
in the background on a smartphone, and collect rich data on phone
use (including data on communications, mobility, social media behavior,
and much more). Digital credit apps use this information to allocate
loans to people who appear creditworthy (i.e., for whom $\hat{y_{i}}$
exceeds some threshold). Since financial regulations prevented us
from actually underwriting loans to research subjects, we instead
focused on analogous problems where a decisionmaker wishes to allocate
resources to individuals with specific characteristics---for instance,
by paying individuals who have a certain income level, or other characteristic
(e.g., intelligence, level of activity, education).\footnote{While these target predictions may bear little resemblance to credit-worthiness,
there are many settings where characteristics like these are being
inferred by digital traces (for example, welfare programs that target
unmarried women, or digital advertisers who target college students).} This allows us to focus on the mechanics of manipulation in a prediction
task, which is the same regardless of which outcome is predicted.

\subsubsection*{Smartphone app}

\begin{figure}
\caption{Smart Sensing App\label{fig:App-Screenshots}}

\begin{minipage}[t]{0.3\columnwidth}%
\begin{center}
{\scriptsize{}(a) Installation Screen}{\scriptsize\par}
\par\end{center}
\begin{center}
{\scriptsize{}\includegraphics[height=8cm]{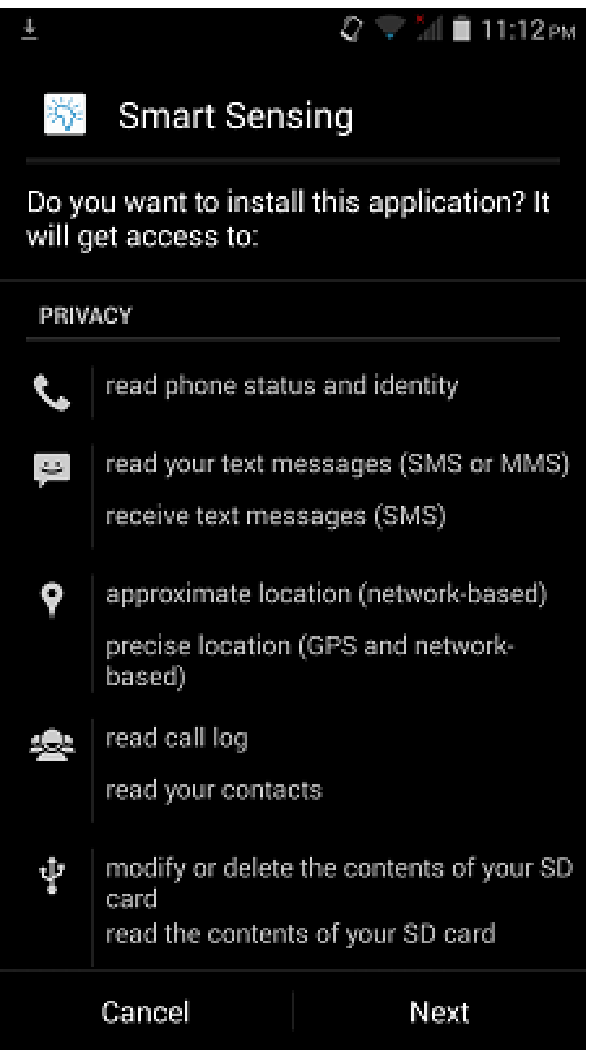}}{\scriptsize\par}
\par\end{center}%
\end{minipage}\hfill{}%
\begin{minipage}[t]{0.3\columnwidth}%
\begin{center}
{\scriptsize{}(b) Challenge with Hint}{\scriptsize\par}
\par\end{center}
\begin{center}
{\scriptsize{}\includegraphics[height=8cm]{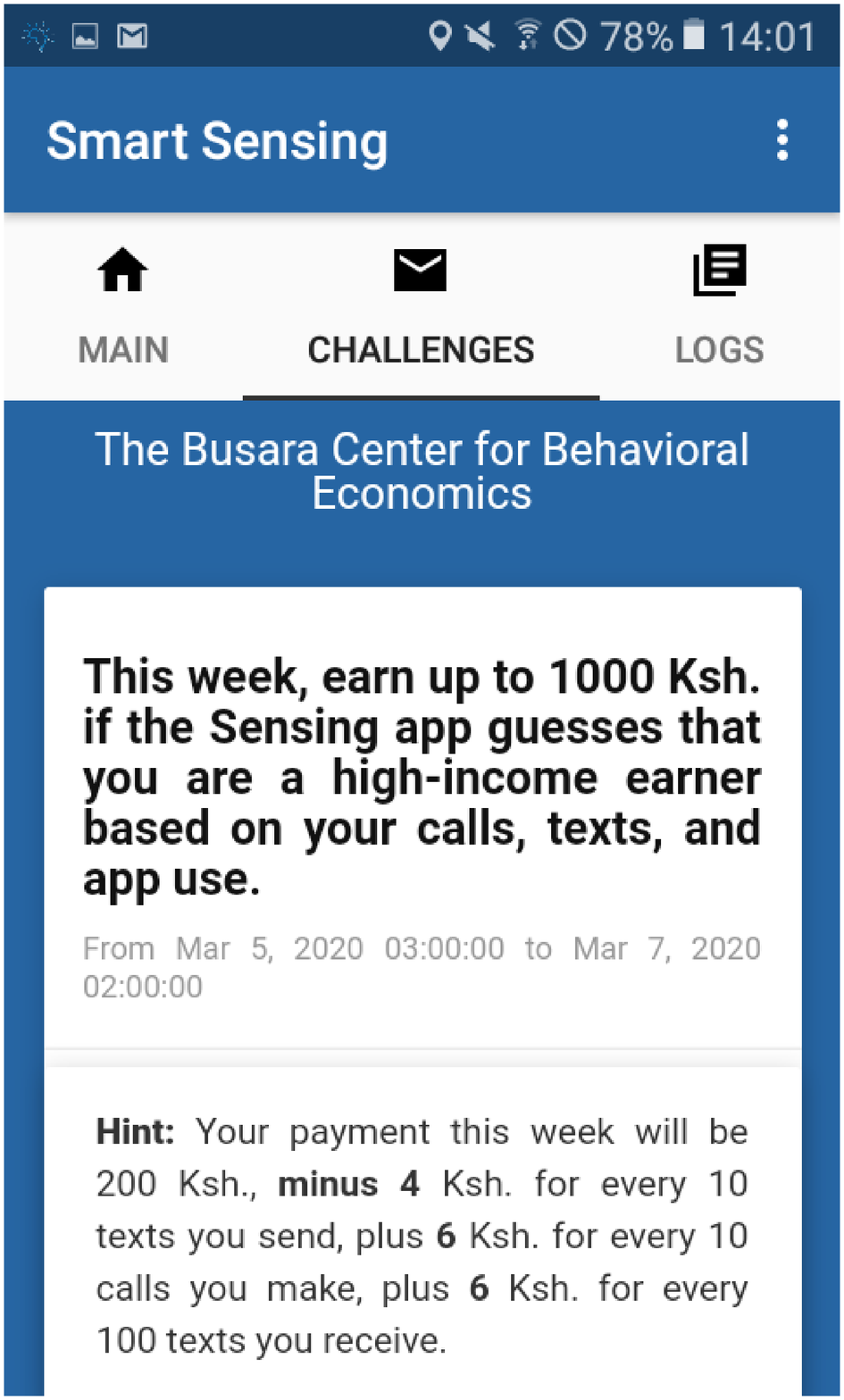}}{\scriptsize\par}
\par\end{center}%
\end{minipage}\hfill{}%
\begin{minipage}[t]{0.3\columnwidth}%
\begin{center}
{\scriptsize{}(c) Earnings Calculator}{\scriptsize\par}
\par\end{center}
\begin{center}
{\scriptsize{}\includegraphics[height=8cm]{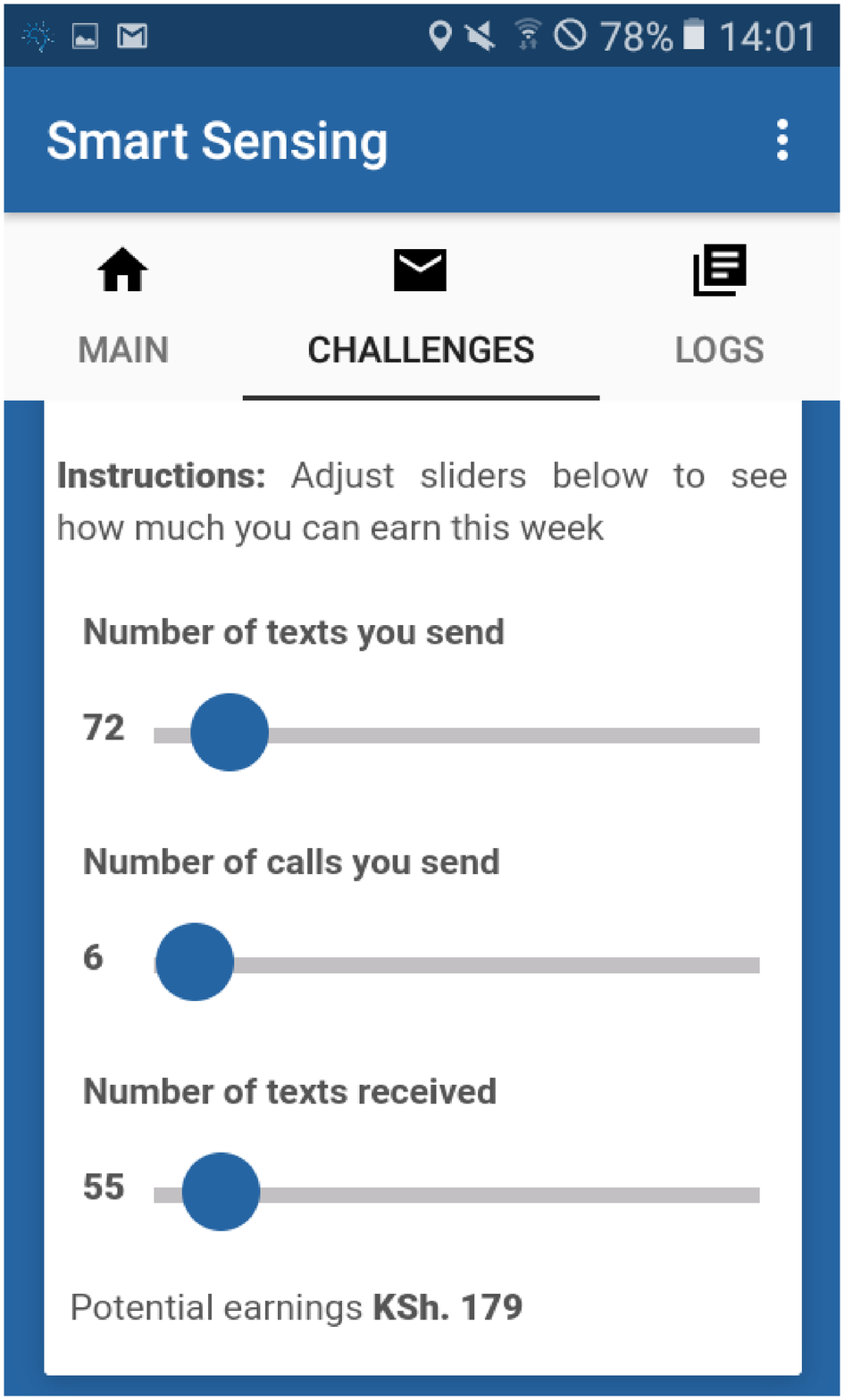}}{\scriptsize\par}
\par\end{center}%
\end{minipage}

\footnotesize{}
\end{figure}

The `Smart Sensing' app we built has  has two key features. First,
it runs in the background on the smartphone to capture anonymized
metadata on how individuals use their phones, such as when calls or
texts are placed, which apps are installed and used, geolocation,
battery usage, wifi connections, and when the screen was on. In total,
we extract over $\bar{K}>1,000$ behavioral features --- Appendix
Figure \ref{fig:Correlation-between-Behaviors} shows the correlation
between 80 different behavioral indicators (``features'') collected
through the app.\footnote{The app is designed to capture this data with minimal impact on battery
life and performance. Data is uploaded to secure Busara servers at
a set frequency, or can be uploaded manually.} Second, the app provides a platform to deliver weekly ``challenges''
to research subjects (see Figure \ref{fig:App-Screenshots}). These
challenges appear on the subject's phone, and offer financial incentives
based on their behavior. The challenges can be very simple (`You will
receive 12 Ksh. for every incoming call you receive this week') or
more complex (`Earn up to 1000 Ksh. if the Sensing app guesses you
are a high-income earner'). Users are paid a base amount of 100 Ksh.
for uploading data, plus any challenge winnings, directly via M-PESA
at the conclusion of each week.

\subsubsection*{Study population and recruitment}

The subject population consists of Kenyans aged 18 years or older
who own a smartphone and are able to travel to the Busara center in
Nairobi. Participants were recruited through in person solicitations
in public spaces in neighborhoods around Nairobi. From this master
list of potential participants, every third individual was saved for
a `top up' sample; we drew invited individuals from this list to participate
later in the experiment, to form a fresh test sample. The remaining
sample was invited at the beginning. All individuals were sequentially
invited for an enrollment session at the Busara center. (The center
had a capacity to enroll 200 people per week.) During enrollment,
participants complete a survey on a tablet on demographics and technology
usage. These responses will form the ground truth about users that
we seek to infer based on phone usage behavior.

Prospective participants were given the opportunity to install the
Sensing App on their phones for about 16 weeks. Participants were
told the dimensions of behavior that would be captured and used anonymously,
and assured that no content of calls or text messages would be recorded.
Participants were given the opportunity to ask questions. Participants
showed understanding of the privacy tradeoffs involved, and voiced
trust in Busara based on its positive reputation in this community.
Participants who opted in to the study were offered help installing
the Sensing App, which provided the main interaction of the study.
During installation, participants had the opportunity to view the
Android permissions required and to decide whether to accept. Our
sample includes only participants who opted in. Participants could
elect to receive challenges in English, Swahili, or both. 82.6\% elected
English, 15.9\% elected Swahili, and 1.4\% elected both. 

\subsubsection*{Weekly rhythm}

The study follows a weekly rhythm. Each Wednesday at noon, each user
receives a generic notification, `Opt in to see this week's challenge!',
via Android notifications and a text message. When a user opens the
app, it will ask them to opt in to a challenge for that week. Only
after a user opts in are the details of their challenge for that week
revealed (see Figure \ref{fig:App-Screenshots}).\footnote{To minimize the possibility of differential attrition, the pre-opt-in
notification was the same for all users regardless of their assigned
challenge.} Challenges are valid until 6pm Tuesday. At the conclusion of the
challenge, users have 16 hours to ensure that their data is uploaded
(until 10am Wednesday). Busara then computes and sends any payments
to users via M-PESA by noon Wednesday, and users receive the next
challenge.

Each week, participants could attrit in two ways: by not uploading
their data, or by not opting in to the challenge.\footnote{As some participants may upload data sparsely throughout the week,
only those who upload within the 21-hour window at the end of the
challenge-week (between 1pm Tuesday and 10am Wednesday) will be counted
as having fully uploaded all of their weekly data.} Participants who failed to upload or opt in were sent text message
reminders, or called by Busara staff, following an attrition protocol
detailed in Appendix \ref{subsec:Attrition-Management}. We include
in our analysis only participant-weeks where the participant opted
in, and uploaded during the end-of-week upload window.

\subsection{Baseline predictions and model estimation \label{sec:ExperimentEstimation}}

\subsubsection*{Predicting user characteristics}

We begin the experiment with baseline weeks that have no incentives
(no active challenges). These baseline weeks allow us to estimate
each individual's type in absence of manipulation, $\boldsymbol{\text{\ensuremath{\underbar{x}}}}$.\footnote{In these `control' weeks, the subject receives a challenge of the
form, `Dear user, you do not have to do anything for this week's challenge.
You will receive an extra Ksh 50 for accepting this challenge.' Our
method could also be used without these control weeks, as long as
there is variation in incentives between weeks; one would then need
to net out the manipulation in estimation.} We estimate each dimension of type using Equation \ref{eq:TypeEst},
with week fixed effects to absorb idiosyncratic weekly shocks.

Consistent with prior work \citep{blumenstock_predicting_2015,bjorkegren_behavior_2019},
we find that characteristics of users can be predicted from phone
behaviors. Results for several outcomes, based on OLS, are shown in
Table \ref{tab:Phase0-OLS}. For characteristics such as monthly income,
intelligence (Ravens Matrices), and overall phone activity, $R^{2}$
values range from 0.02 to 0.15. \emph{} To make these rules easier
for participants to interpret, we will focus on three variable decision
rules selected via LASSO; the last row of Table \ref{tab:Phase0-OLS}
shows that these obtain similar $R^{2}$ when cross validated.

\begin{table}
\caption{Behavior Predicts Individual Characteristics\label{tab:Phase0-OLS}}

\begin{centering}
{\tiny{}}%
\begin{tabular}{lr@{\extracolsep{0pt}.}lr@{\extracolsep{0pt}.}lr@{\extracolsep{0pt}.}lr@{\extracolsep{0pt}.}lr@{\extracolsep{0pt}.}lr@{\extracolsep{0pt}.}l}
 & \multicolumn{4}{c}{} & \multicolumn{8}{c}{}\tabularnewline
\hline 
\hline 
 & \multicolumn{4}{c}{\textbf{\scriptsize{}Monthly Income}} & \multicolumn{4}{c}{\textbf{\scriptsize{}Intelligence}} & \multicolumn{4}{c}{\textbf{\scriptsize{}Activity PCA}}\tabularnewline
{\scriptsize{}OLS} & \multicolumn{2}{c}{} & \multicolumn{2}{c}{} & \multicolumn{4}{c}{{\scriptsize{}(Ravens)}} & \multicolumn{4}{c}{}\tabularnewline
\hline 
{\scriptsize{}Average Duration of Workday Calls} & {\scriptsize{}-6}&{\scriptsize{}877 } & {\scriptsize{}(0}&{\scriptsize{}471)} & {\scriptsize{}0}&{\scriptsize{}0009} & {\scriptsize{}(0}&{\scriptsize{}6)} & {\scriptsize{}-0}&{\scriptsize{}0007} & {\scriptsize{}(0}&{\scriptsize{}185)}\tabularnewline
{\scriptsize{}Average Duration of Outgoing Calls} & {\scriptsize{}5}&{\scriptsize{}746} & {\scriptsize{}(0}&{\scriptsize{}584)} & {\scriptsize{}-0}&{\scriptsize{}0005} & {\scriptsize{}(0}&{\scriptsize{}815)} & {\scriptsize{}0}&{\scriptsize{}0003} & {\scriptsize{}(0}&{\scriptsize{}607)}\tabularnewline
{\scriptsize{}Calls with Non-Contacts} & {\scriptsize{}-27}&{\scriptsize{}747} & {\scriptsize{}(0}&{\scriptsize{}005){*}{*}{*}} & {\scriptsize{}-0}&{\scriptsize{}006} & {\scriptsize{}(0}&{\scriptsize{}001){*}{*}{*}} & {\scriptsize{}0}&{\scriptsize{}0002} & {\scriptsize{}(0}&{\scriptsize{}649)}\tabularnewline
{\scriptsize{}\# Unique Evening Text Contacts} & {\scriptsize{}102}&{\scriptsize{}477} & {\scriptsize{}(0}&{\scriptsize{}129)} & {\scriptsize{}0}&{\scriptsize{}016} & {\scriptsize{}(0}&{\scriptsize{}196)} & {\scriptsize{}0}&{\scriptsize{}003} & {\scriptsize{}(0}&{\scriptsize{}435)}\tabularnewline
{\scriptsize{}Incoming Call Count} & {\scriptsize{}14}&{\scriptsize{}962} & {\scriptsize{}(0}&{\scriptsize{}065)} & {\scriptsize{}0}&{\scriptsize{}001} & {\scriptsize{}(0}&{\scriptsize{}416)} & {\scriptsize{}0}&{\scriptsize{}005} & {\scriptsize{}(0}&{\scriptsize{}0){*}{*}{*}}\tabularnewline
{\scriptsize{}Evening Text Count} & {\scriptsize{}-5}&{\scriptsize{}904} & {\scriptsize{}(0}&{\scriptsize{}194)} & {\scriptsize{}-0}&{\scriptsize{}0007} & {\scriptsize{}(0}&{\scriptsize{}399)} & {\scriptsize{}-0}&{\scriptsize{}0002} & {\scriptsize{}(0}&{\scriptsize{}322)}\tabularnewline
{\scriptsize{}Average Duration of Evening Calls} & {\scriptsize{}-1}&{\scriptsize{}739} & {\scriptsize{}(0}&{\scriptsize{}637)} & {\scriptsize{}0}&{\scriptsize{}0004} & {\scriptsize{}(0}&{\scriptsize{}614)} & {\scriptsize{}0}&{\scriptsize{}0007} & {\scriptsize{}(0}&{\scriptsize{}703)}\tabularnewline
{\scriptsize{}Minimum Duration of Weekend Calls} & {\scriptsize{}2}&{\scriptsize{}950} & {\scriptsize{}(0}&{\scriptsize{}874)} & {\scriptsize{}0}&{\scriptsize{}003} & {\scriptsize{}(0}&{\scriptsize{}406)} & {\scriptsize{}-0}&{\scriptsize{}0008} & {\scriptsize{}(0}&{\scriptsize{}935)}\tabularnewline
{\scriptsize{}Outgoing Texts on Weekdays} & {\scriptsize{}-7}&{\scriptsize{}130} & {\scriptsize{}(0}&{\scriptsize{}417)} & {\scriptsize{}-0}&{\scriptsize{}002} & {\scriptsize{}(0}&{\scriptsize{}225)} & {\scriptsize{}-0}&{\scriptsize{}0001} & {\scriptsize{}(0}&{\scriptsize{}791)}\tabularnewline
{\scriptsize{}Outgoing Text Count} & {\scriptsize{}3}&{\scriptsize{}666} & {\scriptsize{}(0}&{\scriptsize{}621)} & {\scriptsize{}0}&{\scriptsize{}0008} & {\scriptsize{}(0}&{\scriptsize{}585)} & {\scriptsize{}0}&{\scriptsize{}001} & {\scriptsize{}(0}&{\scriptsize{}001){*}{*}{*}}\tabularnewline
{\scriptsize{}Outgoing Call Count} & {\scriptsize{}14}&{\scriptsize{}556} & {\scriptsize{}(0}&{\scriptsize{}004){*}{*}{*}} & {\scriptsize{}-0}&{\scriptsize{}001} & {\scriptsize{}(0}&{\scriptsize{}14)} & {\scriptsize{}0}&{\scriptsize{}004} & {\scriptsize{}(0}&{\scriptsize{}0){*}{*}{*}}\tabularnewline
{\scriptsize{}Incoming Text Count} & {\scriptsize{}1}&{\scriptsize{}762} & {\scriptsize{}(0}&{\scriptsize{}6)} & {\scriptsize{}0}&{\scriptsize{}002} & {\scriptsize{}(0}&{\scriptsize{}013){*}{*}} & {\scriptsize{}0}&{\scriptsize{}001} & {\scriptsize{}(0}&{\scriptsize{}0){*}{*}{*}}\tabularnewline
{\scriptsize{}Intercept} & {\scriptsize{}5259}&{\scriptsize{}547} & {\scriptsize{}(0}&{\scriptsize{}0){*}{*}{*}} & {\scriptsize{}5}&{\scriptsize{}071} & {\scriptsize{}(0}&{\scriptsize{}0){*}{*}{*}} & {\scriptsize{}-0}&{\scriptsize{}956} & {\scriptsize{}(0}&{\scriptsize{}0){*}{*}{*}}\tabularnewline
\hline 
{\scriptsize{}N} & \multicolumn{2}{c}{{\scriptsize{}1539}} & \multicolumn{2}{c}{} & \multicolumn{2}{c}{{\scriptsize{}1557}} & \multicolumn{2}{c}{} & \multicolumn{2}{c}{{\scriptsize{}1415}} & \multicolumn{2}{c}{}\tabularnewline
{\scriptsize{}R2} & {\scriptsize{}0}&{\scriptsize{}0241} & \multicolumn{2}{c}{} & {\scriptsize{}0}&{\scriptsize{}0223} & \multicolumn{2}{c}{} & {\scriptsize{}0}&{\scriptsize{}7593} & \multicolumn{2}{c}{}\tabularnewline
\hline 
 & \multicolumn{2}{c}{} & \multicolumn{2}{c}{} & \multicolumn{2}{c}{} & \multicolumn{2}{c}{} & \multicolumn{2}{c}{} & \multicolumn{2}{c}{}\tabularnewline
{\scriptsize{}OTHER MODELS} & \multicolumn{2}{c}{} & \multicolumn{2}{c}{} & \multicolumn{2}{c}{} & \multicolumn{2}{c}{} & \multicolumn{2}{c}{} & \multicolumn{2}{c}{}\tabularnewline
{\scriptsize{}LASSO: 3 covariate model, 10-fold CV R2} & {\scriptsize{}0}&{\scriptsize{}0180} & \multicolumn{2}{c}{} & {\scriptsize{}0}&{\scriptsize{}0044} & \multicolumn{2}{c}{} & {\scriptsize{}0}&{\scriptsize{}6173} & \multicolumn{2}{c}{}\tabularnewline
\hline 
\end{tabular}{\tiny\par}
\par\end{centering}
\medskip{}

\footnotesize{\textit{Notes}: Each column indicates a different prediction
target. P-values in parentheses. N represents individuals. 10-fold
cross-validated R2 is reported for a LASSO regression where the regularization
parameter is set in order to achieve a 3-covariate model.}
\end{table}

\subsubsection*{Evidence that app-based challenges induce manipulation}

We will eventually use variation in behavior induced by our randomized
experiment to estimate the cost of manipulating different behaviors,
$\mathbf{C}(\mathbf{z}_{i})$. This exogenous variation comes from
weeks when subjects are assigned `simple' challenges that incentivize
modifying a single behavior, of the form, `We\textquoteright ll pay
you $M$ for each additional $x_{j}$ you do', where amount $M$ and
behavior $j$ are assigned randomly. For example, one challenge was,
`You will receive 3 Ksh. for each text you send this week, up to Ksh.
250.' In the long run, individuals may identify new, easier ways
to manipulate these indicators. To mimic this, we held focus groups
to identify the most effective ways to manipulate different features,
and during onboarding, exposed each participant to a discussion of
how one could change different types of behavior (this is similar
to hiring `white hat' hackers to uncover security weaknesses).

People response to these challenges, as anticipated by our theory
(Equation \ref{eq:x(beta)}). For intuition, Table \ref{tab:Phase1Results}
shows how behavior changed in response to simple challenges. Each
column shows a regression of an outcome on different incentives (randomly
assigned). Individuals manipulate the particular behaviors that were
incentivized, as shown by the diagonal, which is positive and significant
for these outcomes. Incentivizing one behavior also affects others,
as shown in the off diagonal elements. For example, incentivizing
missed incoming calls also increases the number of texts sent (presumably
requests to contacts to be called). Our method can theoretically
exploit these cross elasticities.

\begin{table}
\caption{Behavior Changes when Incentivized\label{tab:Phase1Results}}

\begin{centering}
{\scriptsize{}}%
\begin{tabular}{lccccc}
 & \multicolumn{5}{c}{}\tabularnewline
\hline 
\hline 
 & \multicolumn{5}{c}{{\scriptsize{}BEHAVIOR OBSERVED}}\tabularnewline
 & \textbf{\scriptsize{}\# Texts} & \textbf{\scriptsize{}\# Missed} & \textbf{\scriptsize{}\# Missed} & \textbf{\scriptsize{}\# People called} & \textbf{\scriptsize{}\# Calls w non-}\tabularnewline
 & \textbf{\scriptsize{}sent} & \textbf{\scriptsize{}calls} & \textbf{\scriptsize{}calls} & \textbf{\scriptsize{}(workday)} & \textbf{\scriptsize{}contacts (weekend)}\tabularnewline
 &  & {\scriptsize{}(outgoing)} & {\scriptsize{}(incoming)} & {\scriptsize{}(M-F, 9am-5pm)} & \tabularnewline
\hline 
 & \multicolumn{5}{c}{{\scriptsize{}change in actions per  \textcent \hspace{.1cm}  of incentive}}\tabularnewline
\hline 
{\scriptsize{}BEHAVIOR INCENTIVIZED} &  &  &  &  & \tabularnewline
{\scriptsize{}\# Texts sent} & \textbf{\footnotesize{}24.508} & {\footnotesize{}-0.052} & {\footnotesize{}-0.836 } & {\footnotesize{}-0.305 } & {\footnotesize{}-0.022}\tabularnewline
 & {\footnotesize{}(0.0){*}{*}{*}} & {\footnotesize{}(0.929)} & {\footnotesize{}(0.337)} & {\footnotesize{}(0.161)} & {\footnotesize{}(0.953)}\tabularnewline
 &  &  &  &  & \tabularnewline
{\scriptsize{}\# Missed Outgoing Calls} & {\footnotesize{}4.16} & \textbf{\footnotesize{}0.709 } & {\footnotesize{}0.825} & {\footnotesize{}0.128 } & {\footnotesize{}-0.002}\tabularnewline
 & {\footnotesize{}(0.058){*}} & {\footnotesize{}(0.079){*}} & {\footnotesize{}(0.167)} & {\footnotesize{}(0.391)} & {\footnotesize{}(0.995)}\tabularnewline
 &  &  &  &  & \tabularnewline
{\scriptsize{}\# Missed Incoming Calls} & {\footnotesize{}-0.206} & {\footnotesize{}0.324 } & \textbf{\footnotesize{}1.187} & {\footnotesize{}0.22 } & {\footnotesize{}0.502}\tabularnewline
 & {\footnotesize{}(0.942)} & {\footnotesize{}(0.536)} & {\footnotesize{}(0.126)} & {\footnotesize{}(0.255)} & {\footnotesize{}(0.126)}\tabularnewline
 &  &  &  &  & \tabularnewline
{\scriptsize{}\# People Called during Workday} & {\footnotesize{}2.307 } & {\footnotesize{}0.156} & {\footnotesize{}0.68} & \textbf{\footnotesize{}0.497} & {\footnotesize{}0.108}\tabularnewline
 & {\footnotesize{}(0.357)} & {\footnotesize{}(0.734)} & {\footnotesize{}(0.318)} & {\footnotesize{}(0.003){*}{*}{*}} & {\footnotesize{}(0.708)}\tabularnewline
 &  &  &  &  & \tabularnewline
{\scriptsize{}\# Calls w Non-Contacts on Weekend} & {\footnotesize{}-2.022} & {\footnotesize{}-0.056} & {\footnotesize{}1.234 } & {\footnotesize{}0.015} & \textbf{\footnotesize{}1.233}\tabularnewline
 & {\footnotesize{}(0.481)} & {\footnotesize{}(0.916)} & {\footnotesize{}(0.113)} & {\footnotesize{}(0.94)} & {\footnotesize{}(0.0){*}{*}{*}}\tabularnewline
 &  &  &  &  & \tabularnewline
\hline 
{\scriptsize{}Week and Individual Fixed Effects} & {\scriptsize{}X} & {\scriptsize{}X} & {\scriptsize{}X} & {\scriptsize{}X} & {\scriptsize{}X}\tabularnewline
{\scriptsize{}N (person-weeks)} & {\scriptsize{}7976} & {\scriptsize{}7976} & {\scriptsize{}7976} & {\scriptsize{}7976} & {\scriptsize{}7976}\tabularnewline
{\scriptsize{}R2} & {\scriptsize{}0.705} & {\scriptsize{}0.637} & {\scriptsize{}0.552} & {\scriptsize{}0.604} & {\scriptsize{}0.491}\tabularnewline
\hline 
\end{tabular}{\tiny{}}{\tiny\par}
\par\end{centering}
\medskip{}

\footnotesize{\textit{Notes:} P-values in parentheses. Bold indicates
diagonal: effect on behavior $j$ when behavior $j$ is incentivized.
N represents person-weeks when no ``incentive challenge'' was assigned
to the given participant. Individual and weekly fixed effects included,
excluding the first week and first individual hash. Each column represents
a separate regression, over the full set of covariates assigned; only
the first five coefficients reported here. {*} p \textless{}
0.1, {*}{*} p \textless{} 0.05, {*}{*}{*} p \textless{} 0.01.}
\end{table}

Since we have a limited sample on which to estimate costs, our challenges
focus on incentivizing a subset of $K$ focal behaviors (from the
full set of $\bar{K}$). Specifically, we select behaviors $\text{\ensuremath{\underbar{\ensuremath{\mathbf{x}}}}}^{C}$
that are useful in predicting the set of user characteristics that
form the basis for our `complex' challenges. To identify this subset,
we run LASSO regressions for each $\mathbf{y}$ to induce variable
selection, and include the selected variables $\left\{ \text{\ensuremath{\underbar{x}}}_{k}\left|\beta_{k}^{naive}\ne0\right.\right\} $.
For each of these variables, we pair an additional behavior that measures
a similar concept but which we anticipate may be differently easy
to manipulate (for example, if a na{\"i}ve regression selects outgoing
calls, we will also include the variable incoming calls).\footnote{We determined ``similar'' behaviors as those that met at least one
of the following conditions: (1) correlated with the primary behavior
with a coefficient of at least 0.75; (2) was a `close cousin' of the
primary behavior, in that it was a different transformation of a similar
underlying behavior (e.g., for `weekly number of late-night calls',
`maximum number of late-night calls in a single day' would be considered
a close cousin); (3) a cross validated LASSO regression that excluded
the principal behavior from the feature set then newly picked out
this variable in its optimal set. From this list of similar behaviors,
we picked alternates based on our intuition of which behaviors would
substitute the best, and which would be the easiest to explain in
a challenge. } Note that by including only a subset of variables, our procedure
implicitly assumes that omitted variables are costless to manipulate
(and therefore should not be included in any decision rule); we will
thus underestimate the performance that could be attained with our
method if costs were fully estimated.\footnote{Note that this procedure will perform poorly if baseline predictiveness
and manipulation cost are highly negatively correlated: in that case
we may omit a behavior $k$ which is less predictive at baseline but
is more predictive in the counterfactual because it is difficult to
manipulate.} In Section \ref{sec:extensions}, we evaluate other potential
methods to lower the expense of measuring manipulation costs.

\subsubsection*{Estimation}

Finally, we use the data from all weeks of the experiment to jointly
estimate types and manipulation costs (using GMM with the moment conditions
outlined in Section \ref{subsec:Manipulation-Costs}). We allow manipulation
cost to differ by behavior, by whether a person reports having high
tech skills, and by an unobserved random effect by person.\footnote{We have allowed for a single dimension of observed heterogeneity in
costs $\mathbf{z}$; with the rest absorbed into unobserved heterogeneity
$V$. Thus Spence signaling will only be captured in that dimension
$\mathbf{z}$. With a larger sample one could estimate a more nuanced
functional form for the observable portion, which would better capture
the correlations between gaming ability $\gamma_{i}$ and bliss behavior
$\text{\ensuremath{\underbar{\ensuremath{\mathbf{x}}}}}_{i}$.} Table \ref{tab:ChallengeResults_Parameters} summarizes these estimated
costs. With our sample size, we find that off diagonal elements are
noisily estimated, so we penalize them to zero ($\lambda_{offdiagonal}^{costs}\to\infty$);
this results in a diagonal cost matrix $\mathbf{C}$.

Several intuitive patterns can be discerned from the estimated manipulation
costs in the top panel of Table \ref{tab:ChallengeResults_Parameters}
(here we present only behaviors selected by models; see Supplemental
Appendix for all estimated costs). Outgoing communications are less
costly to manipulate than incoming communications. Text messages,
which are relatively cheap to send, are more manipulated than calls,
which are relatively expensive. We also find that complex behaviors
(such as the standard deviation of talk time; estimated but not shown
on this summary diagram) are less manipulable than simpler behaviors
(such as the average duration of talk time).

Costs are also heterogeneous across people, as shown in the bottom
panel of Table \ref{tab:ChallengeResults_Parameters}. On average
it is 10\%pt easier for individuals who report advanced or higher
tech skills to manipulate their mobile phone behaviors. Overall, including
unobserved heterogeneity in gaming ability, the 90th percentile finds
it 2.5 times easier to game than the 10th percentile.\emph{}

\begin{table}
\caption{Estimated Manipulation Costs\label{tab:ChallengeResults_Parameters}}

{\footnotesize{}}{\footnotesize\par}
\begin{centering}
{\scriptsize{}}%
\begin{tabular}{lccccc}
 & \multicolumn{3}{c}{} & \tabularnewline
\hline 
\hline 
\multicolumn{6}{l}{\textbf{Heterogeneity by Behavior} (\textbf{$\mathbf{C}$ }diagonal;
subset of behaviors selected by models)}\tabularnewline
\multicolumn{6}{l}{\includegraphics[bb=0bp 0bp 800bp 400bp,width=1\columnwidth]{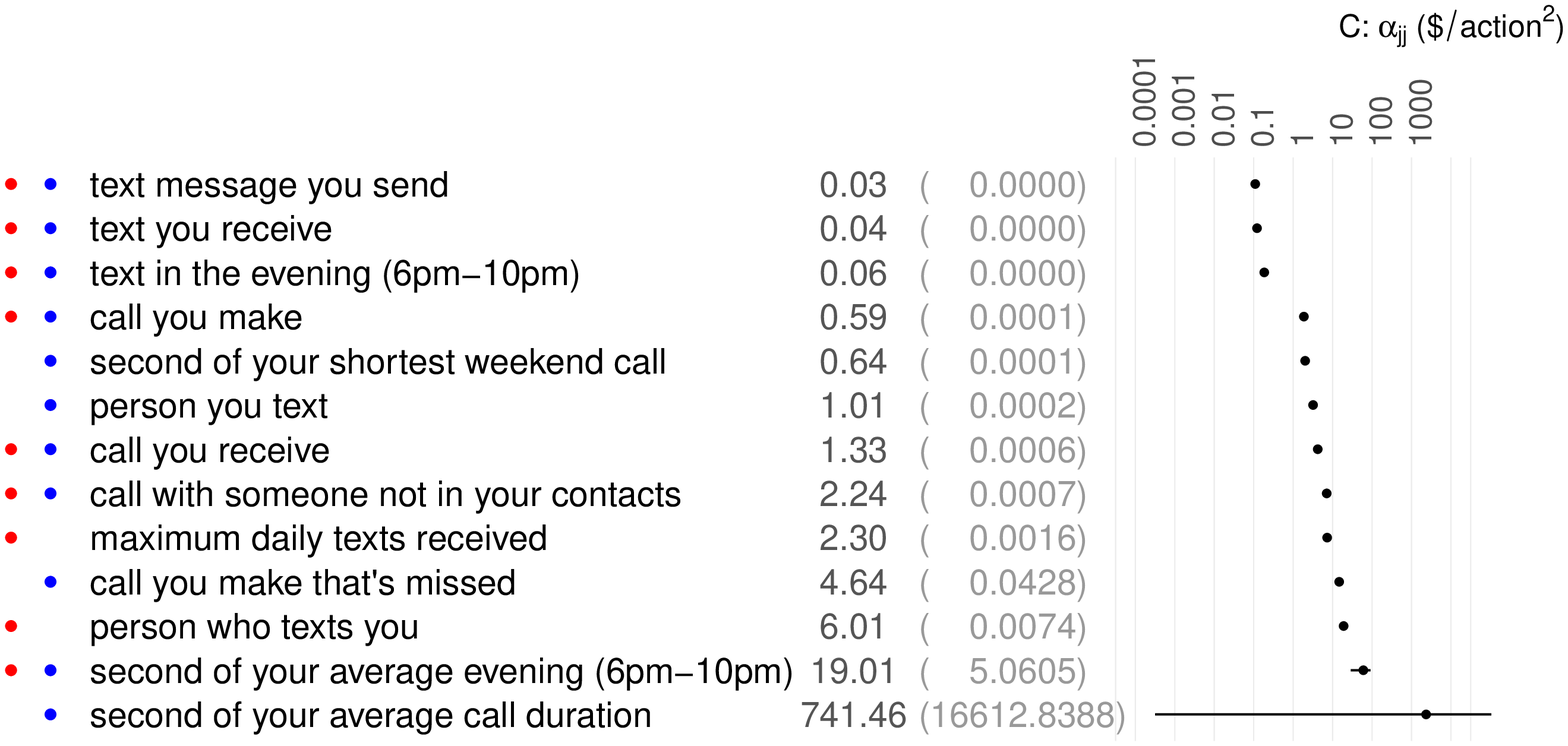}}\tabularnewline
 &  &  &  &  & \tabularnewline
 &  &  &  &  & \tabularnewline
\hline 
\multicolumn{6}{l}{\textbf{Heterogeneity by Person }(\textbf{$\gamma_{i}$})}\tabularnewline
 &  &  &  &  & \tabularnewline
$\gamma_{i}$ & $=$ & \multicolumn{2}{c}{$e^{-\omega z_{i}}$} & + & \multicolumn{1}{c}{$v_{i}$}\tabularnewline
 &  & {\small{}Low tech skills} & {\small{}1.00} &  & \multirow{3}{*}{\includegraphics[height=2cm]{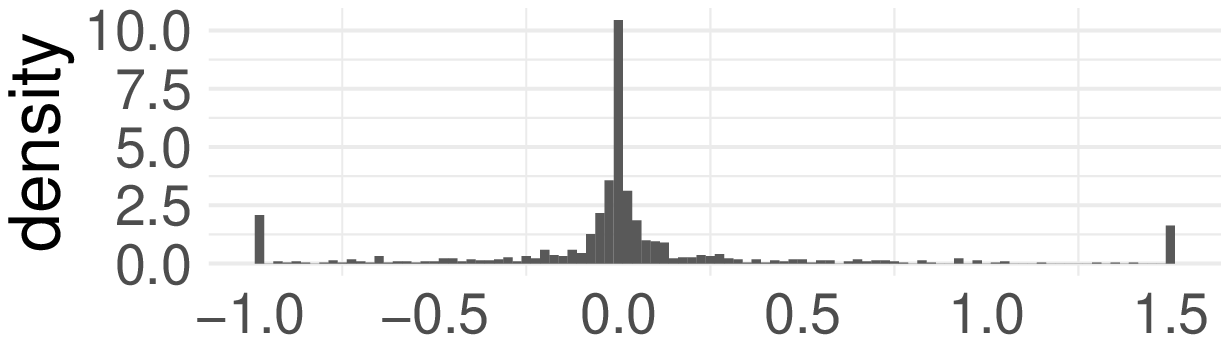}}\tabularnewline
 &  & {\small{}High tech skills} & {\small{}1.10} &  & \tabularnewline
 &  &  &  &  & \tabularnewline
 &  &  &  &  & \tabularnewline
\hline 
\end{tabular}{\scriptsize\par}
\par\end{centering}
\medskip{}

{\footnotesize{}\footnotesize{In top panel: Red: used in a LASSO
model; blue: used in SR model. Line segment represents standard error.
Parameters estimated using GMM. In cost matrix, off diagonal elements
$\alpha_{jk};\text{j\ensuremath{\ne k}}$ regularized to zero ($\lambda_{offdiagonal}^{costs}\to\infty$),
diagonal elements regularized with $\lambda_{diagonal}^{costs}=1.0$,
set via cross validation. Standard errors estimated from PD approximation
of inverse Hessian. Shown here with }\textbf{$v_{i}$ }winsorized
at top and bottom of range; in implementation, only bottom is winsorized,
to maintain assumption of non-negative \textbf{$\gamma_{i}$}.{\footnotesize{}
}Only behaviors selected by models shown in Panel I; for all behaviors
see Supplemental Appendix{\footnotesize{}.}}{\footnotesize\par}
\end{table}

\subsection{Results: Naive vs. Robust Decisions}
\label{sec:results}

The final and most important stage of the experiment compares decisions
made by standard machine learning algorithms to the decisions made
by our new strategy-robust estimator that accounts for the cost of
manipulating behavior. The robust decision rules can be directly estimated
with Equation \ref{eq:estimator}, which relies on the estimates of
$\boldsymbol{\text{\ensuremath{\underbar{x}}}}$ and $\mathbf{C}_{i}$
that come from previous stages of the experiment.

In this final stage, subjects receive complex challenges that reward
them for their ultimate classification, of the form `We\textquoteright ll
pay you M if you are classified as $\hat{y}$.' We consider a focal
challenge of the form, `Earn up to 1000 Ksh. if the Sensing app guesses
you are a high-income earner.' These challenges are designed to mimic
real world applications of machine learning, where depending on how
they are classified, users may receive a loan (digital credit), grant
(targeted aid), or other benefits.

\subsubsection*{Estimating Decision Rules}

In order to keep decision rules simple and interpretable for our participants,
we consider decision rules of up to three features. We regularize
na{\"i}ve decision rules to three features, selecting $\lambda^{decision}=\max(\lambda^{cv},\text{\ensuremath{\underbar{\ensuremath{\lambda}}}}^{3var})$,
where $\text{\ensuremath{\underbar{\ensuremath{\lambda}}}}^{3var}$
is the smallest hyperparameter that results in a 3 variable LASSO
model. We use the same hyperparameter to penalize our strategy robust
decision rule, and allow it to select only among three variable models.\footnote{For a given $\lambda^{decision}$ that selects three variables in
a LASSO model, the strategy robust model will tend to select more
than three variables, because it induces some penalization on its
own. Instead of restricting to three variable models, one could alternately
increase $\lambda^{decision}$.}

\subsubsection*{Treatments}

Participants are randomly assigned into different targets ($\hat{y}$),
decision rules (standard: $\boldsymbol{\beta}^{LASSO}$, or robust
$\boldsymbol{\beta}^{stable}$), and whether the decision rule is
kept opaque or revealed transparently to the user. Under the opaque
treatment, users are told only the outcome and the reward. Under the
transparent treatment, users see the coefficients of the decision
rule, which reveals how much they are rewarded for changing which
behaviors. We included an interactive interface that can be used to
compute the payments that would result from different behavior (see
Figure \ref{fig:App-Screenshots}c). Because the transparent treatment
reveals information about potential decision rules, after a person
has seen a transparent challenge for $\hat{y}$, we do not assign
them to an opaque challenge for the same outcome.

Table \ref{tab:Shift-In-Behavior} summarizes the effect of decision
rule incentives on behavior. High income people make more outgoing
calls, and send fewer texts but receive more. If we pay people to
`act like a high-income earner,' without revealing the decision rule,
the response is noisy and often in the wrong direction (participants
place fewer calls and send more texts). Participants who are transparently
presented with the decision rule change their behavior, closer to
the direction incentivized by the algorithm, though the response is
still noisy.

\begin{table}
\caption{Agents Game Algorithms\label{tab:Shift-In-Behavior}}

\begin{centering}
{\scriptsize{}}%
\begin{tabular}{lccccc}
 & \multicolumn{5}{l}{}\tabularnewline
\hline 
\hline 
 & {\scriptsize{}\# Calls} & {\scriptsize{}\# Texts} & {\scriptsize{}\# Texts} & {\scriptsize{}\# Calls w Non-Contacts} & {\scriptsize{}Mean Call Duration}\tabularnewline
 & {\scriptsize{}(outgoing)} & {\scriptsize{}(outgoing)} & {\scriptsize{}(incoming)} & {\scriptsize{}(incoming + outgoing)} & {\scriptsize{}(evening, seconds)}\tabularnewline
\hline 
\multicolumn{6}{l}{\textbf{\scriptsize{}Weekly Challenge:}\textbf{\textit{\scriptsize{}
Use your phone like a high-income earner!}}}\tabularnewline
\textbf{\scriptsize{}Panel I:} & \multicolumn{5}{l}{\textbf{\scriptsize{}Incentives Generated by Algorithm}{\scriptsize{}
(\textcent/action)}}\tabularnewline
{\footnotesize{}$\boldsymbol{\beta}^{LASSO}$} & {\footnotesize{}0.625} & {\footnotesize{}-0.395} & {\footnotesize{}0.065} & {\scriptsize{}0} & {\scriptsize{}0}\tabularnewline
 &  &  &  &  & \tabularnewline
\hline 
\textbf{\scriptsize{}Panel II:}{\scriptsize{} $\mathbf{x}_{it}$} & \multicolumn{5}{l}{}\tabularnewline
{\scriptsize{}Assigned to challenge,} & {\footnotesize{}-6.5573} & {\footnotesize{}14.3701} & {\footnotesize{}12.0135} & {\footnotesize{}1.1672} & {\footnotesize{}-6.8104}\tabularnewline
{\scriptsize{}algorithm opaque} & {\footnotesize{}(9.949)} & {\footnotesize{}(16.405)} & {\footnotesize{}(20.583)} & {\footnotesize{}(3.473)} & {\footnotesize{}(7.002)}\tabularnewline
 &  &  &  &  & \tabularnewline
{\scriptsize{}Assigned to challenge,} & {\footnotesize{}11.8231} & {\footnotesize{}-15.69} & {\footnotesize{}-11.907} & {\footnotesize{}0.6706} & {\footnotesize{}-4.5744}\tabularnewline
{\scriptsize{}algorithm transparent} & {\footnotesize{}(9.083)} & {\footnotesize{}(14.976)} & {\footnotesize{}(18.79)} & {\footnotesize{}(3.17)} & {\footnotesize{}(6.392)}\tabularnewline
 &  &  &  &  & \tabularnewline
\hline 
{\scriptsize{}N (Person-weeks)} & {\scriptsize{}1664} & {\scriptsize{}1664} & {\scriptsize{}1664} & {\scriptsize{}1664} & {\scriptsize{}1664}\tabularnewline
\hline 
\end{tabular}{\scriptsize\par}
\par\end{centering}
\medskip{}

\footnotesize{\textit{Notes:} The first panel reports the decision
rule associated with the challenge. The second reports the results
of a regression of behavior on challenge assignment. Regressions estimated
based on dummy indicators for complex challenge assignment for participants
assigned ``income'' challenge, over person-weeks when the income
challenge was assigned or when no challenge was assigned (``control''
weeks). Simple challenge assignment person-weeks, used in estimating
costs, are not included. Standard errors in parentheses.}
\end{table}

\subsubsection*{Performance of decision rules}

We compare performance of na{\"i}ve vs. robust decision rules in Table
\ref{tab:Phase2}. The first two columns (under `Income') show results
for the challenge that incentivized participants to use their phones
like a high-income earner; the last two columns show the performance
averaged across several different challenges. The decision rules and
associated manipulation costs are shown in the top panel (``Decision
Rule''); the relative performance of the different estimators is shown
below (under ``Prediction Error''). We note several results.

\begin{table}
\caption{Strategy Robust vs. Standard Decision Rules\label{tab:Phase2}}

\begin{centering}
{\footnotesize{}}%
\begin{tabular}{llccccccc}
 &  & \multicolumn{2}{c}{} &  &  &  &  & \tabularnewline
\hline 
\hline 
 &  & \multicolumn{2}{c}{\textbf{\footnotesize{}Income}} &  & \emph{\footnotesize{}Costs} &  & \multicolumn{2}{c}{\textbf{\footnotesize{}All Outcomes (Pooled)}}\tabularnewline
 &  &  &  &  &  &  & \multicolumn{2}{c}{{\scriptsize{}Income, Intelligence, Activity PCA}}\tabularnewline
\textbf{\footnotesize{}Decision Rule} &  & {\footnotesize{}$\boldsymbol{\beta}^{LASSO}$} & {\footnotesize{}$\boldsymbol{\beta}^{stable}$} &  & {\footnotesize{}$\text{\ensuremath{\alpha_{jj}}}$} &  &  & \tabularnewline
 &  & \multicolumn{2}{c}{{\scriptsize{}\textcent/action}} &  & {\scriptsize{}\textcent/action}\textsuperscript{{\scriptsize{}2}} &  &  & \tabularnewline
\cline{1-4} \cline{2-4} \cline{3-4} \cline{4-4} \cline{6-6} \cline{8-9} \cline{9-9} 
{\footnotesize{}\# Calls (outgoing)} &  & {\footnotesize{}0.625} & {\footnotesize{}0.542} &  & {\footnotesize{}0.591} &  & {\footnotesize{}.} & {\footnotesize{}.}\tabularnewline
{\footnotesize{}\# Texts (outgoing)} &  & {\footnotesize{}-0.395} & {\footnotesize{}-0.107} &  & {\footnotesize{}0.035} &  & {\footnotesize{}.} & {\footnotesize{}.}\tabularnewline
{\footnotesize{}\# Texts (incoming)} &  & {\footnotesize{}0.065} & {\footnotesize{}0} &  & {\footnotesize{}0.038} &  & {\footnotesize{}.} & {\footnotesize{}.}\tabularnewline
{\footnotesize{}\# Texts (6pm-10pm)} &  & {\footnotesize{}0} & {\footnotesize{}-0.121} &  & {\footnotesize{}0.058} &  & {\footnotesize{}.} & {\footnotesize{}.}\tabularnewline
 &  &  &  &  &  &  &  & \tabularnewline
\cline{1-4} \cline{2-4} \cline{3-4} \cline{4-4} \cline{6-6} \cline{8-9} \cline{9-9} 
\textbf{\footnotesize{}Prediction Error} &  & \multicolumn{2}{c}{{\scriptsize{}RMSE (\$)}} &  &  &  & \multicolumn{2}{c}{{\scriptsize{}RMSE (\$)}}\tabularnewline
{\footnotesize{}Baseline Data: Control} &  & {\footnotesize{}3.55} & {\footnotesize{}3.55} &  &  &  & {\footnotesize{}3.70} & {\footnotesize{}3.75}\tabularnewline
{\footnotesize{}Baseline Data: Predicted Transparent} &  & {\footnotesize{}4.66} & {\footnotesize{}3.83} &  &  &  & {\footnotesize{}4.34} & {\footnotesize{}3.85}\tabularnewline
 &  &  &  &  &  &  &  & \tabularnewline
{\footnotesize{}Implemented: Opaque} &  & {\footnotesize{}3.24} & {\footnotesize{}3.23} &  &  &  & {\footnotesize{}4.00} & {\footnotesize{}3.80}\tabularnewline
{\footnotesize{}Implemented: Transparent} &  & {\footnotesize{}3.87} & {\footnotesize{}3.66} &  &  &  & {\footnotesize{}4.93} & {\footnotesize{}4.31}\tabularnewline
 &  &  &  &  &  &  &  & \tabularnewline
{\footnotesize{}Predicted Cost of Transparency} &  &  & {\footnotesize{}$\le$0.28} &  &  &  &  & {\footnotesize{}$\le$0.15}\tabularnewline
{\footnotesize{}Equilibrium Cost of Transparency} &  &  & {\footnotesize{}$\le$0.41} &  &  &  &  & {\footnotesize{}$\le$0.31}\tabularnewline
 &  &  &  &  &  &  &  & \tabularnewline
\hline 
{\footnotesize{}Average Payout (\$)} &  & {\footnotesize{}3.30} & {\footnotesize{}3.24} &  &  &  & {\footnotesize{}3.23} & {\footnotesize{}2.98}\tabularnewline
 &  &  &  &  &  &  &  & \tabularnewline
{\footnotesize{}N (Control Person-Weeks)} &  & {\footnotesize{}3781} & {\footnotesize{}3781} &  &  &  & {\footnotesize{}3781} & {\footnotesize{}3781}\tabularnewline
{\footnotesize{}N (Treatment Person-Weeks, Opaque)} &  & {\footnotesize{}85} & {\footnotesize{}85} &  &  &  & {\footnotesize{}230} & {\footnotesize{}230}\tabularnewline
{\footnotesize{}N (Treatment Person-Weeks, Trans.)} &  & {\footnotesize{}91} & {\footnotesize{}74} &  &  &  & {\footnotesize{}252} & {\footnotesize{}216}\tabularnewline
\hline 
\end{tabular}{\tiny{}}{\tiny\par}
\par\end{centering}
\smallskip{}

\footnotesize{\textit{Notes:} The first panel reports the decision
rule associated with the challenge, and the costs associated with
these behaviors. The second reports the performance of the different
models over the groups they were assigned to; on the left, the naive
LASSO regression, and on the right, this paper's strategy-robust (SR)
model. Performance figures estimated using a regression of model indicators
on week-model RMSE, weighted by number of person-weeks. `Transparent
Predicted' RMSE denotes the RMSE that our theoretical model expected,
given costs of manipulation and behavioral incentives. `Predicted
Cost of Transparency' denotes the difference between predicted transparent
RMSE under the SR model and baseline RMSE under the naive LASSO. `Equilibrium
Cost of Transparency' denotes the difference between implemented transparent
SR model RMSE and opaque naive model RMSE. Pooled performance is estimated
using this same regression approach, after combining all model-weeks
over the three outcomes investigated: a PCA of phone activity, intelligence,
and monthly income. Full regression results and standard errors reported
in appendix.}

\end{table}

First, in the top panel, we observe important differences in the decision
rules estimated by $\beta^{LASSO}$ vs. $\beta^{act}$. LASSO places
weight on the behaviors that were most correlated at baseline: outgoing
calls, outgoing texts, and incoming texts. However, the estimated
costs of manipulating some of these behavior -- and in particular
the costs of manipulating text messaging behavior -- are low, and
therefore likely to be manipulated when incentivized. Thus, our strategy
robust decision rule both selects less manipulable behaviors (evening
texts rather than incoming texts), and shrinks manipulable behaviors
(especially outgoing texts).

We evaluate prediction error using root mean squared error (RMSE),
in units of dollars, in the middle panel. The magnitude of error is
similar to the average payout, around \$3 for a week. The first row
shows prediction error in the baseline data: LASSO performs slightly
better than our strategy robust estimator when no manipulation is
expected. But when people manipulate their behavior, our method is
expected to perform better, as shown in the second row.

When actually implemented, our method performs better when the decision
rule is transparent (average error \$3.66 instead of \$3.87 for income;
or \$4.31 vs. \$4.93 for all outcomes pooled). When the decision rule
is opaque, we find that our method performs comparably to or slightly
better than LASSO, possibly due to increased shrinkage (\$3.23 vs.
\$3.24 for income; \$3.80 vs. \$4.00 for all outcomes pooled). Table
\ref{tab:Perf} reports results for all outcomes.

Even if a policymaker intended to keep the decision rule opaque, using
our robust method can reduce systematic risk in the chance that agents
discover the decision rule. In practical implementations, policymakers
could adaptively tweak the level of robustness to match the level
of manipulation. An ad hoc approach could select a convex combination
of the naive and robust models; a more nuanced approach could model
consumers' uncertainty about the model.

\subsubsection*{Cost of transparency}

Our framework provides a way to bound a key cost of imposing algorithmic
transparency \citep{akyol_price_2016}. Many tech firms argue that
imposing transparency would reduce the quality of machine decisions,
because rules may perform better if they can rely on opacity to prevent
manipulation. Our method allows us to bound this performance cost.
We can compare the performance arising from the optimal opaque rule
(under the assumption that opacity will prevent it from being manipulated)
to the optimal equilibrium transparent rule (factoring in equilibrium
manipulation). Because the opaque rule also faces the threat of manipulation,
this difference is the upper bound of the performance cost of imposing
transparency, arising from increased manipulation.

The most straightforward way to measure this cost of transparency
would require disclosing the decision rule to a subset of users, and
assessing any drop in performance after a process of equilibration.
But for the most consequential decisions, once the decision rule is
revealed to some, it can leak out to the entire market. Such disclosure
irreversibly tips the market to transparency, and thus is a nonstarter
for policy discussions.

Crucially, under the assumptions of our model, this quantity can be
estimated without revealing the decision rule: it only requires the
estimation of types and costs (the first part of our experiment).\footnote{Our method of estimating costs does requires revealing the existence
of features to users, but does not require specifying whether those
features are included in the model, or with what weights (one could
estimate costs for a large set of features, hiding the features critical
to the model).} Our method makes it possible for regulators or firms to assess the
cost that transparency would impose---prior to making their model
transparent. Our model based estimates suggest that transparency introduces
a performance cost of $\le\$0.28$ (8\% of baseline error) for our
income targeting rule, or $\le\$0.15$ (4\%) for all outcomes pooled
together. These numbers are shown in the final rows of the middle
panel of Table \ref{tab:Phase2}.

When we actually implement transparency in our experiment, we find
that the performance cost is similar to these model based estimates:
$\le\$0.41$ (13\%) for income, or $\le\$0.31$ (8\%) for all outcomes
pooled together. (To mitigate the problem of leakage, we only assess
opaque performance prior to each individual observing a transparent
challenge for that outcome. Because our decision rules were not going
to be used later in production, we were unconcerned about them leaking
out after the experiment.)

\section{Extensions}
\label{sec:extensions}

\subsection{Alternate methods to estimate manipulation costs}

Our method requires estimating $C$ and $\gamma_{i}$, which are new
objects. The experimental approach we use is likely not feasible in
many settings. We offer suggestions on alternative approaches to measure
these costs.

\begin{figure}
\caption{Expert Elicited Manipulation Cost Estimates\label{fig:Expert-Costs}}
\medskip{}

\begin{centering}
\begin{tabular}{>{\raggedright}p{4.4cm}>{\raggedright}p{4.4cm}}
\multicolumn{1}{c}{\textbf{(a)} Reduced Form Shift in Behavior} & \multicolumn{1}{c}{\textbf{(b)} Structural Cost Estimates}\tabularnewline
\includegraphics[height=5cm]{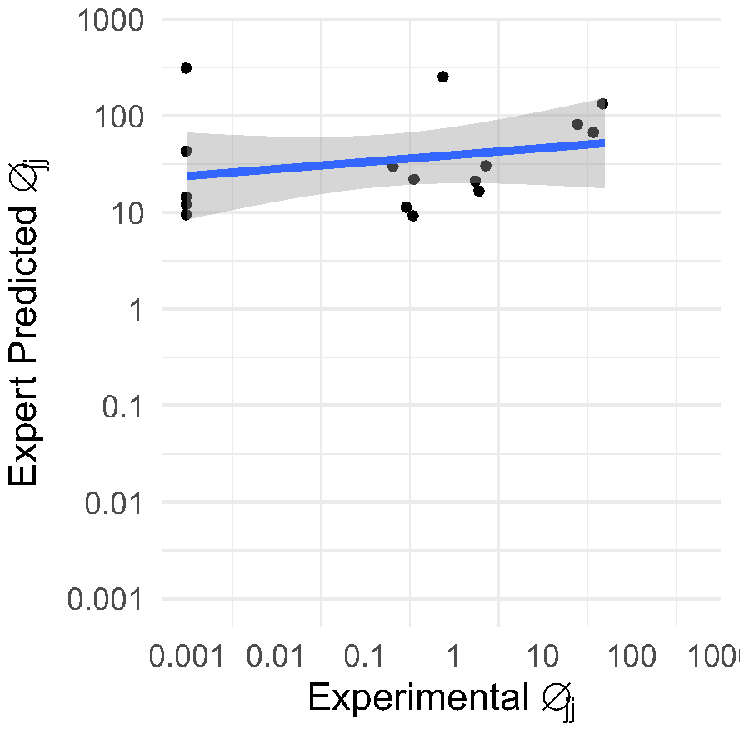} & \includegraphics[height=5cm]{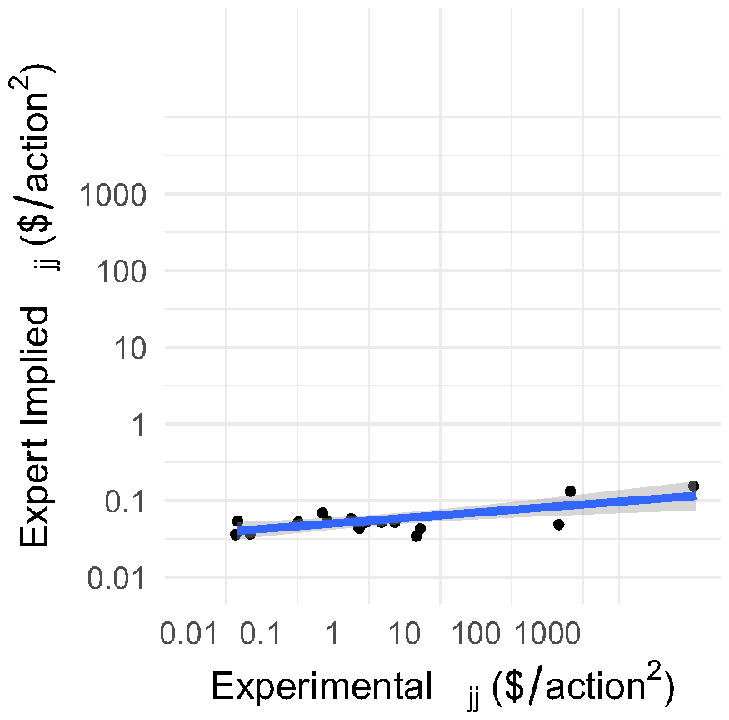}\tabularnewline
\end{tabular}
\par\end{centering}

\footnotesize{For structural costs we set $\bar{\gamma}=1$ and $\hat{\alpha}_{jj}=\frac{\bar{\gamma}\cdot\beta_{j}}{\max(0.001,\Delta_{jj})}$.}
\end{figure}

\emph{Expert elicitations. }We evaluate how well experts can predict
the costs of manipulating different behaviors, using a method similar
to \citet{dellavigna_predicting_2016}. We sent a survey to 177 experts
with different backgrounds (PhDs from different fields, research assistants,
Busara staff who had not worked on the experiment, and Mechanical
Turk workers in the US) to predict how Kenyans would manipulate different
phone behaviors when incentivized. Results are shown in Figure \ref{fig:Expert-Costs}.
In panel A, we compare the predicted change in behavior from a given
incentive to the actual experimental estimate ($\Delta_{jj}\coloneqq x_{j}(\beta_{j})-\bar{x}_{j}$).
In Panel B, we compare the implied structural cost estimates (for
predicted costs $\hat{\alpha}_{jj}=\frac{\bar{\gamma}\cdot\beta_{j}}{\Delta_{jj}}$);
although experts predict that costs are too low, the correlation is
0.75. This suggests that it may be possible to use expert elicitations
to estimate manipulation costs.

\emph{Partially estimated.} The costs of behavior $k$ may be related
to that of behavior $k'$. Because of this, we may be able to predict
unknown cost $\alpha_{kj}$ based on correlations between types $\boldsymbol{\text{\ensuremath{\underbar{x}}}}$
and known costs, for some prediction function: $\hat{\alpha}_{kj}=f(C,\boldsymbol{\text{\ensuremath{\underbar{x}}}})$.

\subsection{Nonlinear decision rules}

To sharpen intuition, this paper focuses on linear decision rules.
While many modern machine learned decision rules are nonlinear, agents'
beliefs about those rules may be well approximated by linear functions.
In such a context, our derivations could be viewed as linear approximations
to both these beliefs, and the actual functions. Additionally, it
may be that some benefits of extreme nonlinearities that can surface
in modern machine learning are lessened when manipulation is taken
into account: contract theory suggests that linear decision rules
are more robust \citep{holmstrom_aggregation_1987,carroll_robustness_2015}.\footnote{With the exception that linear models can be subject to the influence
of outliers; one may thus want to tamp down inputs as they approach
the boundaries of the distribution of training data.} 

Our approach could also be extended to work in nonlinear settings.
In nonlinear environments there may also be multiple equilibria. In
such a setting, if iterative learning converges, it may converge to
a local optimum, whereas an approach like ours could be used to select
a global equilibrium.\footnote{Thanks to Glen Weyl for this point.}

\section{Conclusion}
\label{sec:conclusion}

This paper considers the possibility that the implementation of machine
decisions changes the world they describe. We focus on the case where
individuals manipulate their behavior in order to game decision rules.
Our chief contribution is to derive decision rules that anticipate
this manipulation, by embedding a behavioral model of how individuals
will respond. This structural approach makes it possible to decompose
decision rules into constituent components, and to gather data on
how those components can be manipulated. From these components, our
structural model allows us to understand how \emph{any} proposed decision
rule of a given form would be manipulated. This allows us to compute
decision rules that are optimal in equilibrium.

We demonstrate our method in a field experiment in Kenya, by deploying
a tailor-made smartphone app that mimics the `digital credit' loan
products that are now commonplace in sub-Saharan Africa. We find that
even some of the world's poorest users of technology -- who are relatively
recent adopters of smartphones and to whom whom the concept of an `algorithm'
is quite foreign \citep{musya_how_2018} -- are savvy enough to change
their behavior to game machine decisions. In this setting, we show
that our strategy robust estimator outperforms standard estimators
on average by 13\%  when individuals are given information about
the scoring rule. This framework also allows us to quantify the ``cost
of transparency'', i.e., the loss in predictive performance associated
with moving from ``security through obscurity'' (with a naive decision
rule) to a regime of full algorithmic transparency (with our strategy-robust
rule). We estimate this loss to be roughly 8\% in equilibrium --
substantially less than the 23\% loss associated with making the naive
rule transparent.

Our discussion focuses on the simple case of linear models with a
small number of predictor variables, where subjects have either no
information or full transparency of the scoring rule. We envision
useful extensions to more complex models and more nuanced beliefs.
More generally, our approach of embedding a model of behavior within
a machine learning estimator may be relevant to a wide range of contexts
where machine learning systems face a changing human environment.
In this sense, it offers a machine learning interpretation of \citet{lucas_econometric_1976},
where algorithmic decisions change the context of the systems they
model. For example, financial forecasts may affect the underlying
financial processes they attempt to describe, personalized news
recommendations may change the information seeking behaviors of consumers,
and predictions about the intensity of a disease may affect individuals' protective 
behaviors and thus its realized intensity.

\clearpage
\begin{spacing}{1}
\bibliographystyle{aer}
\bibliography{manipulation}
\end{spacing}

\clearpage
\newpage

\section*{\center{Appendices}}
 \renewcommand{\thetable}{A\arabic{table}}
\renewcommand{\thefigure}{A\arabic{figure}}
\renewcommand{\thesection}{A\arabic{section}}
\setcounter{table}{0}
\setcounter{figure}{0}
 \setcounter{section}{0}

\section{Experimental Design}

\subsection{Pre Analysis Plan}

This study was pre-registered with the AEA RCT Registry (AEARCTR-0004649)
prior to the experiment (September 3, 2019).\footnote{Prior to the collection of the main outcomes in phase 2, we amended
the registration, adding one sentence that specifies that the focal
performance measure will be mean squared error (which corresponds
with the objective minimized by the method; January 15, 2020). We
later noticed that the registration still contained text in another
section that appeared to specify that the focal measures would be
$R^{2}$ or AUC; prior to the completion of phase 2 and prior to analysis
of the main outcomes, we amended the registration to delete that sentence
(February 4, 2020).}

Our implementation deviated in several respects from the pre analysis
plan: at the start of phase 2 the cloud server account ran out of
storage space, and the Busara center was hit by a power outage due
to construction on a nearby road. These two events disrupted servers
for several hours during the upload window, and caused some participants'
phones to become overloaded with records. It took several weeks to
recover the affected participants. Because of the disruption, we extended
phase 2 and delayed the expert cost surveys.

\subsection{Attrition Management\label{subsec:Attrition-Management}}

Attrition in the context of this study had two dimensions: first,
there were participants who do not regularly upload data through
their app, and second, there were participants who did not participate
in the assigned weekly challenges. (As some participants may have uploaded
data sparsely throughout the week, only those who uploaded within the
21-hour window at the end of the challenge-week {[}between 1pm Tuesday
and 10am Wednesday{]} were counted as having fully uploaded all
of their weekly data.)

In order to minimize both such types of attrition, participants were 
sent regular reminders via text to encourage engagement. Every
participant in the study was sent a text every Tuesday at 1pm
to remind them to upload their data through the Smart Sensing app.

Additionally, on Wednesday, Thursday and Friday, participants who still had
not uploaded data or activated their challenge respectively were
contacted by phone and surveyed by the Busara team. Specifically,
the protocol was as follows:
\begin{itemize}
\item On Wednesday, participants who had not uploaded any data during the
five day period ending on Wednesday at 12pm were contacted and
surveyed, as were those who uploaded some data in this period but
not during the `end-of-week upload window' (between 6pm Tuesday and
10am Wednesday)
\item On Thursday, participants whose phones showed that they did not receive
a challenge by Thursday 12pm were contacted and surveyed, as were
participants whose phones show that they did receive a challenge but
who had not opted in to accept the challenge.
\item On Friday, participants whose phones showed that they still had not
received and opted-in to a challenge were contacted and surveyed, as were
participants whose phones showed that they did receive a challenge but
who had not opted in to accept the challenge.
\end{itemize}
For all of the above categories, any participant who did not answer
a call on the first attempt would be re-contacted once
more by the surveyor after the rest of the calls were complete.

Finally, to mitigate the effects of attrition during the analysis
stage, any participant-weeks wherein the participant did not opt in
and/or did not upload during the end-of-week upload window were
dropped from the sample prior to all analysis. During baseline weeks,
a single passive challenge was assigned to all participants, offering
a flat bonus to upload data within the upload window; in this way,
we ensured that our analysis control groups would also be restricted
to those who opt in to this passive challenge, and were thus a valid
comparison group to the restricted panel during the challenge weeks.

\subsection{Communicating Decision Rules}

In focus groups we found that individuals had difficulty understanding
decimals or complicated mathematical operations (e.g., standard deviation).
We stuck to simple behaviors and formatted decision rules as follows,
to make it easier for participants to understand how their marginal
behavior affects their payment:
\begin{itemize}
\item Each coefficient is rounded to the nearest integer. If the nearest
integer is zero, the denominator was inflated by factors of 10 until
it became nonzero. (If the unit was seconds or minutes, the denominator
was instead inflated by factors of 60.)
\item The order of indicators was randomized between three orderings (ABC,
CAB, BCA for indicators A, B, and C).
\item The constant term was reported last, unless the first coefficient
was negative, in which case the constant was reported first.
\end{itemize}

\clearpage
\newpage
\section{Appendix Figures}
\label{app:figures}


\begin{figure}[hb]
\caption{Comparative Statics\label{fig:Comparative-Statics-Simple}}
\centering{}%
\renewcommand{\arraystretch}{1.2}
\begin{tabular}{>{\raggedright}p{7.5cm}>{\raggedright}p{7.5cm}}
\multicolumn{2}{>{\raggedright}p{15cm}}{\emph{Note:} The first behavior is more predictive in the baseline behavior ($b_{1}>b_{2}$), but is easily manipulable ($\alpha_{11}\ll\alpha_{22}$). Below panels show weights on coefficients as manipulation costs are scaled for:} \\[1cm]
\multicolumn{1}{c}{\textbf{(a) $x_{2}$}} & \multicolumn{1}{c}{\textbf{(b) Interaction $x_{1}$, $x_{2}$ }}\\
\multicolumn{1}{c}{$\boldsymbol{\beta}^{stable}(\alpha_{22})$} & \multicolumn{1}{c}{$\boldsymbol{\beta}^{stable}(\alpha_{12})$} \\
\multicolumn{1}{c}{$\alpha_{12}=0$} & \multicolumn{1}{c}{$\alpha_{22}=32$}  \\
\includegraphics[height=4.5cm]{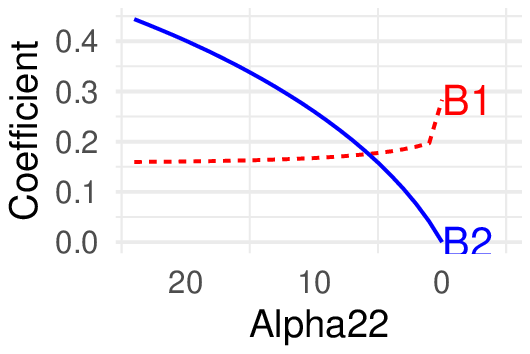} & 
\includegraphics[height=4.5cm]{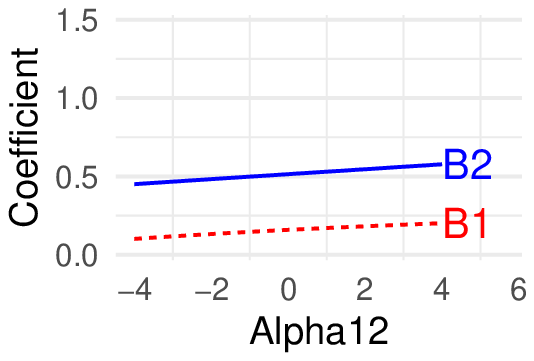} \tabularnewline
\footnotesize{As $x_{2}$ becomes cheaper to manipulate ($\alpha_{22}$ decreases), $\boldsymbol{\beta}^{stable}$ places less weight on it, and adjusts the weight placed on $x_{1}$.} &
\footnotesize{If manipulating one variable makes it easier to manipulate the other ($\alpha_{12}$ sufficiently negative), $\boldsymbol{\beta}^{stable}$ reduces weight on both.}\tabularnewline 
& \\ \hline
& \\
\multicolumn{2}{>{\raggedright}p{15cm}}{\footnotesize{$\boldsymbol{\text{\ensuremath{\underbar{x}}}}_{i}\overset{iid}{\sim}N\left(0,1\right)$,
$\mathbf{b}=[1.4,1]$, $\mathbf{C}=\frac{1}{\gamma\gamma_{i}}\left[\begin{array}{cc}
4 & \alpha_{12}\\
\alpha_{12} & \text{\ensuremath{\alpha_{22}}}
\end{array}\right]$, $\frac{1}{\gamma_{i}}\overset{iid}{\sim}Uniform\left[0,10\right]$,
$\epsilon_{i}\overset{iid}{\sim}N\left(0,0.25\right)$. Squared error
measured on an out of sample draw from the same population, incentivized
to that decision rule.}}\tabularnewline
\end{tabular}
\end{figure}

\begin{sidewaysfigure}
\caption{\textbf{Additional Comparative Statics\label{fig:Comparative-Statics-Additional}}}
\medskip{}

\begin{centering}
\begin{tabular}{>{\raggedright}p{4.4cm}>{\raggedright}p{4.4cm}>{\raggedright}p{0.5cm}|>{\raggedright}p{0.5cm}>{\raggedright}p{4.4cm}>{\raggedright}p{4.4cm}}
\multicolumn{2}{>{\raggedright}p{8.8cm}}{\textbf{$C$ and $\gamma_{i}$ heterogeneous}

The first behavior is more predictive in the baseline behavior ($b_{1}>b_{2}$),
but is easily manipulable ($\alpha_{11}\ll\alpha_{22}$).} &  &  & \textbf{$C$ homogenous: }Features equally costly to manipulate & \textbf{$\gamma_{i}$ homogeneous: }Same gaming ability\tabularnewline
\multicolumn{1}{c}{} & \multicolumn{1}{c}{} &  &  &  & \tabularnewline
\multicolumn{1}{c}{$\boldsymbol{\beta}^{Ridge}(\lambda;\gamma\textnormal{ fixed})$} & \multicolumn{1}{c}{$\boldsymbol{\beta}^{stable}(\lambda|\mathbf{\gamma}=1)$} &  &  & \multicolumn{1}{c}{\textbf{(d)} $\boldsymbol{\beta}^{stable}(\nicefrac{1}{\gamma})$} & \multicolumn{1}{c}{\textbf{(e)} $\boldsymbol{\beta}^{stable}(\nicefrac{1}{\gamma})$}\tabularnewline
\includegraphics[width=4cm]{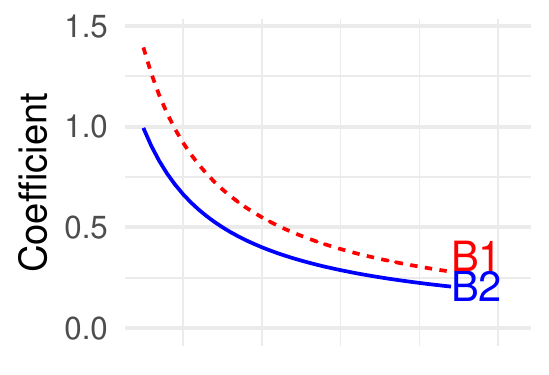} &
 \includegraphics[width=4cm]{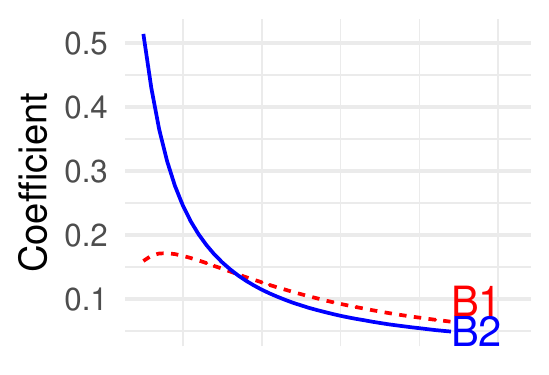} &
  &  & \includegraphics[height=3cm]{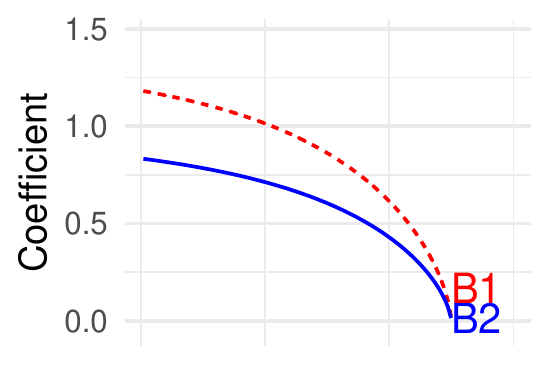} &
 \includegraphics[height=3cm]{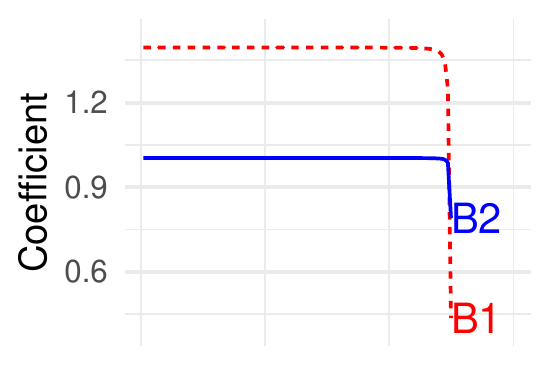}\tabularnewline
\includegraphics[width=4cm]{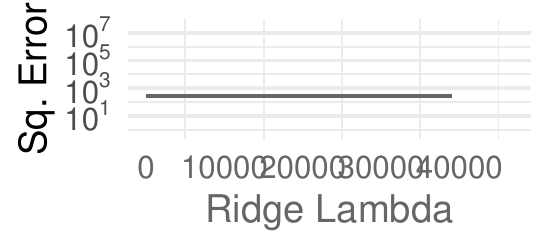} &
 \includegraphics[width=4cm]{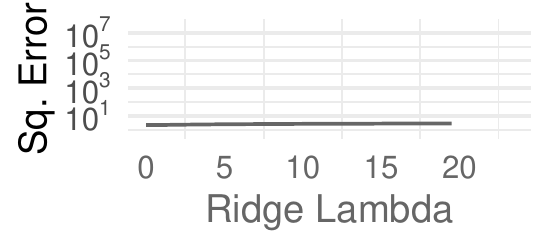} &
  &  & \includegraphics[height=2cm]{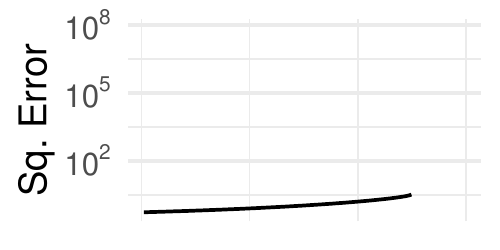} & 
\includegraphics[height=2cm]{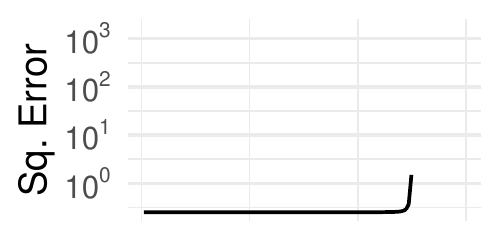}\tabularnewline
 &  &  &  & \includegraphics[height=2cm]{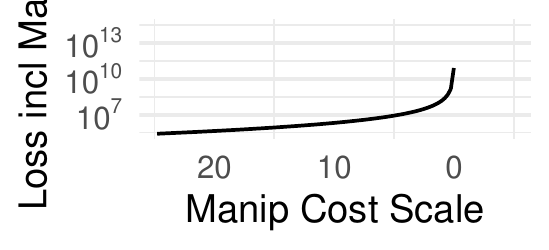} &
 \includegraphics[height=2cm]{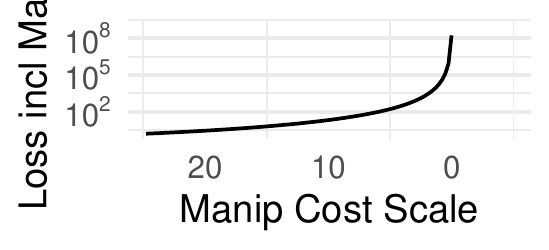}\tabularnewline
 & \centering{}{\scriptsize{}Manipulation Cost = 1} &  &  &  & \tabularnewline
\footnotesize{Like LASSO, ridge places more weight on $x_{1}$.} & \footnotesize{Our method can be combined with other forms of penalization
(such as ridge shown here), to more finely manage out of sample fit.} &  &  & \footnotesize{When features are equally costly to manipulate, our
method penalizes in a similar manner to ridge.} & \footnotesize{When gaming ability is homogenous, everyone shifts
behavior equally. Predictive performance remains high, but utility
is wasted on manipulation.}\tabularnewline
 &  & \multicolumn{1}{>{\raggedright}p{0.5cm}}{} &  &  & \tabularnewline
\multicolumn{6}{>{\raggedright}p{22cm}}{\footnotesize{$\boldsymbol{\text{\ensuremath{\underbar{x}}}}_{i}\overset{iid}{\sim}N\left(0,1\right)$,
$\mathbf{b}=[1.4,1]$, $\mathbf{C}^{het}=\frac{1}{\gamma\gamma_{i}}\left[\begin{array}{cc}
4 & 0\\
0 & 32
\end{array}\right]$, $\mathbf{C}^{hom}=\frac{1}{\gamma\gamma_{i}}\left[\begin{array}{cc}
8 & 0\\
0 & 8
\end{array}\right]$, $\frac{1}{\gamma_{i}^{het}}\overset{iid}{\sim}Uniform\left[0,10\right]$,
$\gamma_{i}^{hom}=5$, $e_{i}\overset{iid}{\sim}N\left(0,0.25\right)$.
Squared error measured on an out of sample draw from the same population,
incentivized to that decision rule.}

}\tabularnewline
\end{tabular}
\par\end{centering}
\end{sidewaysfigure}

\begin{center}
\begin{figure}
\caption{\label{fig:Correlation-between-Behaviors}}

\begin{centering}
\includegraphics[width=0.8\columnwidth]{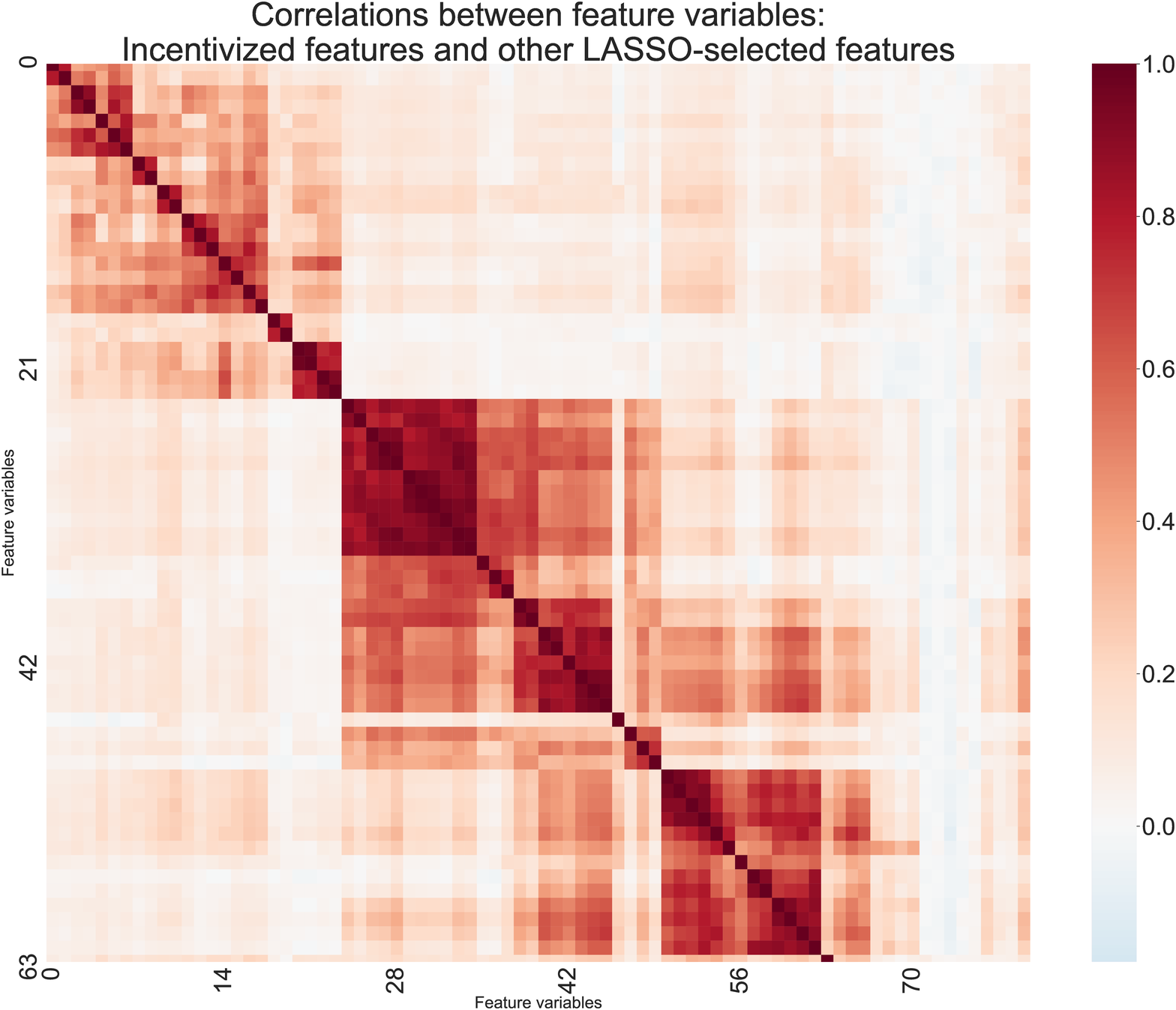}
\par\end{centering}
\footnotesize{Each row and column represent a feature of behavior.
Features are clustered into similar groups. The diagonal indicates
that the correlation of a feature with itself is +1.}
\end{figure}
\par\end{center}


\clearpage
\newpage
\section{Appendix Tables}
\label{app:tables}

\input{tables/mc-industry}

\input{tables/mc-signal}

\input{tables/performance}


\end{document}

%% file: tables/mc-basic.tex
\begin{table}[p]
\centering
\caption{Manipulation Can Harm Prediction (Monte Carlo)\label{tab:Monte-Carlo-Simulation-NonEquilibrating}}
\renewcommand{\arraystretch}{1.3}
\begin{threeparttable}
\begin{tabular}{lccccccc}
\toprule
\multicolumn{5}{c}{\footnotesize\textbf{Decision Rule}}  & & \multicolumn{2}{c}{\footnotesize\textbf{Performance} (squared loss)}  \\
 &  {$\beta_{0}$} & {$\beta_{1}$} & {$\beta_{2}$} & {$\beta_{3}$} &  & {No manip.} & {Manipulation}\\
\cline{1-5} \cline{7-8}
\multicolumn{8}{l}{\footnotesize\emph{Panel A:} \textbf{Data generating process}} \\
\cline{1-5} \cline{7-8}
{$\mathbf{\boldsymbol{b}}^{DGP}$}  & {0.200} & {3.000} & {0.100} & {0.100} &  & {0.267} & {3745.046}\\[.8em]
\multicolumn{8}{l}{\footnotesize\emph{Panel B:} \textbf{Standard Approaches}} \\
\cline{1-5} \cline{7-8}
{$\boldsymbol{\beta}^{OLS}$}  & {0.205} & {3.042} & {0.061} & {0.116} &  & {0.266} & {3961.225}\\[.5em]
\multicolumn{8}{l}{\footnotesize \hspace{.25cm} \emph{`Industry' Approach (estimated cumulatively)}}\\
{$\boldsymbol{\beta}^{OLS(1)}$ after $\boldsymbol{\beta}^{OLS}$}  & {-0.798} & {0.061} & {2.090} & {-1.675} &  & {3.275} & {625.762}\\
{$\boldsymbol{\beta}^{OLS(2)}$ after $\boldsymbol{\beta}^{OLS}$} & {-2.174} & {0.174} & {0.436} & {0.143} &  & {12.861} & {8.369}\\
{$\boldsymbol{\beta}^{OLS(3)}$ after $\boldsymbol{\beta}^{OLS}$} & {-1.376} & {0.165} & {0.573} & {0.483} &  & {9.343} & {4.415}\\
{$\vdots$}  &  &  &  &  &  &  & \\
{$\boldsymbol{\beta}^{OLS(100)}$ after $\boldsymbol{\beta}^{OLS}$}  & {-1.619} & {0.316} & {0.753} & {-0.059} &  & {8.442} & {2.105}\\
{$\vdots$}  &  &  &  &  &  &  & \\
{$\boldsymbol{\beta}^{OLS(1000)}$ after $\boldsymbol{\beta}^{OLS}$}  & {-1.854} & {0.489} & {0.582} & {-0.124} &  & {9.211} & {1.959}\\[.8em]
\multicolumn{8}{l}{\footnotesize\emph{Panel C:} \textbf{Strategy Robust Method}} \\
\cline{1-5} \cline{7-8}
{$\boldsymbol{\beta}^{stable}$}  & {-1.813} & {0.503} & {0.536} & {-0.096} &  & {9.155} & {1.939}\\[.5em]
\multicolumn{8}{l}{\footnotesize \hspace{.25cm} \emph{If costs are misestimated:}}\\
{$\boldsymbol{\beta}_{\hat{C}_{i}=2diag(C_{i})}^{stable}$} &  {-1.566} & {0.658} & {0.719} & {-0.352} &  & {6.893} & {10.826}\\[.5em]
\multicolumn{8}{l}{\footnotesize \hspace{.25cm} \emph{Followed by Industry Approach (estimated cumulatively):}}\\
{$\;\;\;$$\boldsymbol{\beta}^{OLS(1)}$ after $\boldsymbol{\beta}_{\hat{C}_{i}=2diag(C_{i})}^{stable}$}  & {-2.045} & {0.800} & {0.042} & {0.418} &  & {10.891} & {4.447}\\
\emph{$\;\;\;$}{$\boldsymbol{\beta}^{OLS(2)}$ after $\boldsymbol{\beta}_{\hat{C}_{i}=2diag(C_{i})}^{stable}$}  & {-2.022} & {0.558} & {0.327} & {0.137} &  & {10.685} & {2.453}\\
\bottomrule
\end{tabular}
\begin{tablenotes}[normal,flushleft]
\footnotesize
\item \emph{Notes}: Monte Carlo simulation results. Panel A shows the coefficients that relate the outcome $(y$) to behaviors ($\mathbf{x}$) under the data generating process (DGP). Panel B shows coefficients from OLS; Panel C shows coefficients estimated with the strategy robust method. Performance is assessed on the same sample of individuals,
under behavior without manipulation: $\mathbf{x}_{i}(\boldsymbol{0})$, or with: $\mathbf{x}_{i}(\boldsymbol{\beta})$. Parameters:\\[.3em]
{\footnotesize
$C=\left[\begin{array}{ccc}
1.0 & 0.1 & 0.2\\
0.1 & 2.0 & 0.8\\
0.2 & 0.8 & 4.0
\end{array}\right]$ , $\boldsymbol{\text{\ensuremath{\underbar{x}}}}\overset{iid}{\sim}N\left(\boldsymbol{0},\left[\begin{array}{ccc}
1 & 1 & 0.1\\
1 & 2 & 1\\
0.1 & 1 & 1
\end{array}\right]\right)$, $\gamma_{i}=\begin{cases}
1 & \text{\ensuremath{\underbar{x}}}_{i1}\le0.2\\
10 & \text{\ensuremath{\underbar{x}}}_{i1}>0.2
\end{cases}$ , $e_{i}\overset{iid}{\sim}N(0,0.25)$}\\
\end{tablenotes}
\end{threeparttable}
\end{table}

%% file: tables/mc-industry.tex
\begin{table}[hbp]
\centering
\caption{Manipulation Can Harm Prediction (Monte Carlo): ``Industry Approach'' \label{tab:Monte-Carlo-Simulation-NonEquilibratingNonCum}}
\renewcommand{\arraystretch}{1.3}
\begin{threeparttable}
\begin{tabular}{lcccccccc}
\toprule
\multicolumn{6}{c}{\footnotesize\textbf{Decision Rule}}  & & \multicolumn{2}{c}{\footnotesize\textbf{Performance} (squared loss)}  \\
 &  & {$\beta_{0}$} & {$\beta_{1}$} & {$\beta_{2}$} & {$\beta_{3}$} &  & {No manip.} & {Manipulation}\\
\cline{1-6}  \cline{8-9} 
\multicolumn{9}{l}{\footnotesize\emph{Panel A:} \textbf{Data generating process}} \\
{$\mathbf{\boldsymbol{b}}^{DGP}$} &  & {0.200} & {3.000} & {0.100} & {0.100} &  & {0.267} & {3745.046}\\[.8em]
\multicolumn{9}{l}{\footnotesize\emph{Panel B:} \textbf{Standard Approaches}} \\
\cline{1-6}  \cline{8-9} 
{$\boldsymbol{\beta}^{OLS}$} &  & {0.205} & {3.042} & {0.061} & {0.116} &  & {0.266} & {3961.225}\\
\multicolumn{9}{l}{\footnotesize \hspace{.25cm} \emph{`Industry' Approach (estimated with just data from that period)}}\\
{$\boldsymbol{\beta}^{OLS(1)}$ after $\boldsymbol{\beta}^{OLS}$} &  & {-0.798} & {0.061} & {2.090} & {-1.675} &  & {3.275} & {625.762}\\
{$\boldsymbol{\beta}^{OLS(2)}$ after $\boldsymbol{\beta}^{OLS}$} &  & {0.172} & {3.111} & {-0.040} & {0.215} &  & {0.270} & {4332.208}\\
{$\boldsymbol{\beta}^{OLS(3)}$ after $\boldsymbol{\beta}^{OLS}$} &  & {-0.755} & {0.120} & {2.077} & {-1.671} &  & {3.071} & {619.059}\\
{$\vdots$} &  &  &  &  &  &  &  & \\
{$\boldsymbol{\beta}^{OLS(1000)}$ after $\boldsymbol{\beta}^{OLS}$} &  & {-0.393} & {3.741} & {-1.341} & {1.566} &  & {1.375} & {11611.884}\\
{$\boldsymbol{\beta}^{OLS(1001)}$ after $\boldsymbol{\beta}^{OLS}$} &  & {-0.404} & {0.704} & {1.861} & {-1.526} &  & {1.674} & {565.383} \\
\bottomrule
\end{tabular}
\begin{tablenotes}[normal,flushleft]
\footnotesize
\item \emph{Notes}: Monte Carlo simulation results. Panel A shows the coefficients that relate the outcome $(y$) to behaviors ($\mathbf{x}$) under the data generating process (DGP). Panel B shows coefficients from OLS, under behavior without manipulation: $\mathbf{x}_{i}(\boldsymbol{0})$, or with manipulation: $\mathbf{x}_{i}(\boldsymbol{\beta})$. Parameters:\\[.3em]
$C=\left[\begin{array}{ccc}
1.0 & 0.1 & 0.2\\
0.1 & 2.0 & 0.8\\
0.2 & 0.8 & 4.0
\end{array}\right]$ , 
$\boldsymbol{\text{\ensuremath{\underbar{x}}}}\overset{iid}{\sim}N\left(\boldsymbol{0},\left[\begin{array}{ccc}
1 & 1 & 0.1\\
1 & 2 & 1\\
0.1 & 1 & 1
\end{array}\right]\right)$ , 
$\gamma_{i}=\begin{cases}
1 & \text{\ensuremath{\underbar{x}}}_{i1}\le0.2\\
10 & \text{\ensuremath{\underbar{x}}}_{i1}>0.2
\end{cases}$ ,
$e_{i}\overset{iid}{\sim}N(0,0.25)$
\end{tablenotes}
\end{threeparttable}
\end{table}

%% file: tables/mc-signal.tex
\begin{table}[p]
\centering
\caption{Manipulation Can Improve Prediction (Monte Carlo)\label{tab:Monte-Carlo-Simulation-ManipulationSignal}}
\renewcommand{\arraystretch}{1.5}
\begin{threeparttable}
\begin{tabular}{p{2cm}cccp{.5cm}cc}
\toprule
\multicolumn{4}{c}{\footnotesize\textbf{Decision Rule}}  & & \multicolumn{2}{c}{\footnotesize\textbf{Performance} (squared loss)}  \\
& $\beta_{0}$ & $\beta_{1}$ & $\beta_{2}$ & &  {No manipulation} & {Manipulation}\\
\cline{1-4} \cline{6-7} 
\multicolumn{7}{l}{\footnotesize\emph{Panel A:} \textbf{Data generating process}} \\
{$\mathbf{\boldsymbol{b}}^{DGP}$} &  {1.00} & {0.10} & {0.01} & & {8.749} & {8.748}\\[.8em]
\multicolumn{7}{l}{\footnotesize\emph{Panel B:} \textbf{Standard Approach}} \\
{$\boldsymbol{\beta}^{OLS}$} &  {1.014} & {-0.003} & {0.130} & &  {8.724} & {8.720}\\[.8em]
\multicolumn{7}{l}{\footnotesize\emph{Panel C:} \textbf{Strategy Robust Method}} \\
{$\boldsymbol{\beta}^{stable}$} &  {1.014} & {-0.022} & {0.156} & & {8.725} & {8.719}\\
\bottomrule
\end{tabular}
\begin{tablenotes}[normal,flushleft]
\footnotesize
\item \emph{Notes}: Monte Carlo simulation results. Panel A shows the coefficients that relate the outcome $(y$) to behaviors ($\mathbf{x}$) under the data generating process (DGP). Panel B shows estimated coefficients from OLS; Panel C shows coefficients estimated with the strategy
robust method. Performance is assessed on the same sample of individuals, under behavior without manipulation: $\mathbf{x}_{i}(\boldsymbol{0})$,
or with: $\mathbf{x}_{i}(\boldsymbol{\beta})$. Parameters:\\
{\footnotesize{}$C=\left[\begin{array}{cc}
2 & 0.5\\
0.5 & 1
\end{array}\right]$} , $\boldsymbol{\text{\ensuremath{\underbar{x}}}}_{i}\overset{iid}{\sim}N\left(\boldsymbol{0},\left[\begin{array}{cc}
2 & 0.5\\
0.5 & 1
\end{array}\right]\right)$, 
$\begin{array}{cl}
\gamma_{i} & =0.1u_{i}-\epsilon_{i}^{3}+B\\
u_{i} & \sim N(0,1)\\
B & \textnormal{set so }\min\gamma_{i}=0.1
\end{array}$ , $\epsilon_{i}\overset{iid}{\sim}N(0,9)$
\end{tablenotes}
\end{threeparttable}
\end{table}

%% file: tables/performance.tex
\begin{sidewaystable}
\centering
\caption{Performance of Decision Rules\label{tab:Perf}}
\renewcommand{\arraystretch}{1.1}
\begin{threeparttable}
{\scriptsize
\begin{tabular}{lllllcccccccc}
\toprule
 & {Costs} & \multicolumn{2}{l}{\textbf{All outcomes (pooled)}} &  & \multicolumn{2}{c}{{Income}} &  & \multicolumn{2}{c}{{Ravens (intelligence)}} &  & \multicolumn{2}{c}{{Activity PCA}} \\
 &  &  &  &  &  &  &  & \multicolumn{2}{c}{{above median}} &  &  &  \\
 & {$\alpha_{jj}$} &  &  &  & {$\boldsymbol{\beta}^{LASSO}$} & {$\boldsymbol{\beta}^{stable}$} &  & {$\boldsymbol{\beta}^{LASSO}$} & {$\boldsymbol{\beta}^{stable}$} &  & {$\boldsymbol{\beta}^{LASSO}$} & {$\boldsymbol{\beta}^{stable}$} \\
 & {¢/action}\textsuperscript{{2}} &  &  &  & \multicolumn{2}{c}{{¢/action}} &  & \multicolumn{2}{c}{{¢/action}} &  & \multicolumn{2}{c}{{¢/action}} \\
\hline 
\textbf{\uline{Decision Rule}} &  &  &  &  &  &  &  &  &  &  &  &  \\
{call\_count\_out} & {0.591} & {-} & {-} &  & {0.625} & {0.542} &  &  &  &  & {1.978} & {1.241} \\
{text\_count\_incoming} & {0.038} & {-} & {-} &  & {0.065} &  &  & {0.278} & {0.145} &  & {1.154} & {0.306} \\
{text\_count\_out} & {0.035} & {-} & {-} &  & {-0.395} & {-0.107} &  &  &  &  &  &  \\
{text\_count\_evening} & {0.058} & {-} & {-} &  &  & {-0.121} &  &  &  &  & {0.036} & {0.35} \\
{calls\_noncontacts} & {2.24} & {-} & {-} &  &  &  &  & {-0.606} & {-0.575} &  &  &  \\
{call\_count\_outgoing\_missed} & {4.64} & {-} & {-} &  &  &  &  & {-0.208} &  &  &  &  \\
{max\_daily\_texts\_incoming} & {2.30} & {-} & {-} &  &  &  &  &  & {0.324} &  &  &  \\
\hline 
\multicolumn{13}{l}{\textbf{\uline{Prediction Error}}}  \\
\textbf{Baseline Data:} &  & \multicolumn{2}{c}{{RMSE (\$)}} &  & \multicolumn{2}{c}{{RMSE (\$)}} &  & \multicolumn{2}{c}{{RMSE (\$)}} &  & \multicolumn{2}{c}{{RMSE (\$)}} \\
{Control} &  & \textbf{3.698} & \textbf{3.745} &  & {3.553} & {3.554} &  & {5.144} & {5.158} &  & {2.396} & {2.523} \\
 &  & {(0.2)} & {(0.19)} &  & {(0.032)} & {(0.032)} &  & {(0.024)} & {(0.025)} &  & {(0.142)} & {(0.076)} \\
{Predicted Transparent} &  & \textbf{4.344} & \textbf{3.85} &  & {4.663} & {3.831} &  & {5.148} & {5.119} &  & {3.319} & {2.592} \\
 &  & {(0.479)} & {(0.622)} &  & {(0.0)} & {(0.0)} &  & {(0.0)} & {(0.0)} &  & {(0.0)} & {(0.0)} \\
\textbf{Implemented:} &  &  &  &  &  &  &  &  &  &  &  &  \\
{Opaque} &  & \textbf{4.002} & \textbf{3.803} &  & {3.243} & {3.232} &  & {5.147} & {5.165} &  & {3.655} & {2.974} \\
 &  & {(0.512)} & {(0.588)} &  & {(0.0)} & {(0.0)} &  & {(0.0)} & {(0.0)} &  & {(0.0)} & {(0.0)} \\
{Transparent} &  & \textbf{4.933} & \textbf{4.308} &  & {3.867} & {3.655} &  & {5.323} & {5.138} &  & {5.762} & {4.014} \\
 &  & {(0.505)} & {(0.41)} &  & {(0.0)} & {(0.0)} &  & {(0.0)} & {(0.0)} &  & {(0.0)} & {(0.0)} \\
 &  &  &  &  &  &  &  &  &  &  &  &  \\
\hline 
\textbf{Cost of Transparency} &  & \textbf{0.931 (0.719)} & \textbf{0.504 (0.717)} &  & {0.624 (0.0)} & {0.423 (0.0)} &  & {0.176 (0.0)} & {-0.027 (0.0)} &  & {2.108 (0.0)} & {1.04 (0.0)} \\
{Eqm.: Predicted} &  & \textbf{0.152 (0.654)} &  &  & {0.278 (0.032)} &  &  & {-0.025 (0.024)} &  &  & {0.196 (0.142)} &  \\
{Eqm.: Implemented} &  & \textbf{0.305 (0.656)} &  &  & {0.412 (0.0)} &  &  & {-0.009 (0.0)} &  &  & {0.36 (0.0)} &  \\
 &  &  &  &  &  &  &  &  &  &  &  &  \\
{Average Payout (\$)} &  & \textbf{3.226} & \textbf{2.979} &  & {3.295} & {3.239} &  & {3.77} & {3.757} &  & {2.612} & {1.94} \\
{N} &  & {114} &  &  & {38} &  &  & {38} &  &  & {38} &  \\
{N person-weeks} &  & {4476} &  &  & {3979} &  &  & {3983} &  &  & {3951} &  \\
\hline 
\end{tabular}{}}{\tiny\par}
\begin{tablenotes}[normal,flushleft]
\scriptsize
\item \emph{Notes}: The first panel reports the decision rule associated with the challenge, and the costs associated with these behaviors. The below panels report the performance of naive LASSO and our strategy-robust model, by outcome and pooled across outcomes, respectively. Performance metrics estimated using a regression
of model indicators on week-model RMSE, weighted by number of person-weeks. Opaque (Training Weeks) represents the average performance of models in control person-weeks, when no behavior was incentivized. Transparent (Model) represents the average expected performance of models given the theoretical model, behavior incentives and estimated costs. Implemented Opaque represents the average performance of models when assigned without transparency hints. Implemented Transparent represents the average performance of models when assigned with transparency hints. Cost of transparency represents the difference between transparent and opaque RMSE for naive LASSO and strategy-robust (SR) models, respectively. 'Eqm.: Predicted' denotes the difference between predicted transparent RMSE under the SR model and baseline RMSE under the naive LASSO. 'Equilibrium Cost of Transparency' denotes the difference between implemented transparent SR model RMSE and opaque naive model RMSE. Average payouts represents the average payout from the assigned challenges associated with the model. $N$ represents the number of model-weeks that the regression is estimated over, $N$ person-weeks represents the number of person-weeks that these model-weeks include, as well as the sum of the weights used in regression. Costs represent structural costs estimated using above procedure. Standard errors in parentheses, clustered at week-outcome level.
\end{tablenotes}
\end{threeparttable}
\end{sidewaystable}